\documentclass[aps,prx,twocolumn,superscriptaddress,floatfix]{revtex4}
\usepackage{graphicx}
\usepackage{dcolumn}
\usepackage{bm}
\usepackage{stmaryrd}
\usepackage{latexsym}
\usepackage{amssymb}
\usepackage{amsfonts}
\usepackage{amsmath}
\usepackage{fancybox}
\usepackage{color}
\usepackage{verbatim}
\usepackage{physics}
\usepackage{bbold}

\def\be{\begin{equation}}
\def\ee{\end{equation}}
\def\ber{\begin{eqnarray}}
\def\eer{\end{eqnarray}}

\begin{document}
\title{Topological flat bands in rhombohedral tetralayer and multilayer graphene on hexagonal boron nitride moire superlattices}
\author{Youngju Park}
\author{Yeonju Kim}
\affiliation{Department of Physics, University of Seoul, Seoul, 02504, Korea}
\author{Bheema Lingam Chittari}
\affiliation{Department of Physical Sciences, Indian Institute of Science Education and Research Kolkata, Mohanpur 741246, West Bengal, India}
\author{Jeil Jung}
\email{jeiljung@uos.ac.kr}
\affiliation{Department of Physics, University of Seoul, Seoul, 02504, Korea}
\affiliation{Department of Smart Cities, University of Seoul, Seoul 02504, Korea}

\begin{abstract}
We show that rhombohedral four-layer graphene (4LG) nearly aligned with a hexagonal boron nitride (hBN) substrate often develops nearly flat isolated low energy bands with non-zero valley Chern numbers.
The bandwidths of the isolated flatbands are controllable through an electric field and twist angle, becoming as narrow as $\sim$10~meV for interlayer potential differences between top and bottom layers of $|\Delta|\approx 10\sim15$~meV 
and $\theta \sim 0.5^{\circ}$ at the graphene and boron nitride interface. 
The local density of states (LDOS) analysis shows that the nearly flat band states are associated to the non-dimer low energy sublattice sites at the top or bottom graphene layers and their degree of localization in the moire superlattice is strongly gate tunable, exhibiting at times large delocalization despite of the narrow bandwidth.
We verified that the first valence bands' valley Chern numbers are $C^{\nu=\pm1}_{V1} = \pm n$, proportional to layer number for $n$LG/BN systems up to $n = 8$ rhombohedral multilayers. 
\end{abstract}
\maketitle
\section{Introduction} 
Nearly aligned van der Waals 2D layered heterostructures leads to moire superlattices~\cite{deheer1,deheer2,deheer3,deheer4,deheer5,ugeda,ohta,lopes,lopes1,shallcross,shallcross1,
shallcross2,shallcross3,bistritzer,koshino,koshino1,jung2014ab,sanjose,sanjose1,sanjose2,Bistritzerprb,wangzf,schmidt, stephen2018,koshino2,vafek1,vafek,vishwanath,vishwanath2} with 
enlarged moire lattice constants ($\ell_m$) and reduced moire Brillouin zones (MBZ) 
where the bandwidth suppression typically enhances the Coulomb correlation effects.
Twisted bilayer graphene (tBG) is a representative system showing a variety of
ordered phases near its magic angle ($\approx 1^{\circ}$) as a function of carrier doping~\cite{mott1,mott2,mott3,super1,super2,super3}. 
Other related graphene moire systems such as twisted trilayer graphene (tTG)~\cite{tTG,tTG_Shin2021,jmpark2021,hao2021,ramires2021,qin2021,phong2021,kim2022,fischer2022,shen2023,yankowitz2023}, twisted monolayer-bilayer graphene (tMBG)~\cite{tMBG,rademaker2020,tMBG_MA2021,xu2021,chen2021,he2021,li2022,tong2022}, and twisted double bilayer graphene (tDBG)~\cite{tbbg, tbbg1, tbbg2, tbbg3, tbbg4, tbbg5, tbbg6,tbbg7} are current systems of interest. 
The gapped massive Dirac layer-based heterostructures~\cite{levitovsong,srivani2019} including transition metal 
 dichalcogenides homo and hetero twisted bilayers~\cite{mitmanish,mitmanish2,wu2019,zhan2020,wang2020,zhang2020,shabani2021,rademaker2022} 
achieves reduced bandwidths at relatively large twist angles allowing to enhance the Coulomb interactions.
Another interesting 2D layered moire system is formed by graphene on hexagonal boron nitride (G/BN) moire superlattices~\cite{bistritzer,lopes,jung2014ab,dillonwong,abhay,nadj,moon2014} with lattice mismatch~\cite{levitovsong,yankowitz}.
When a graphene layer ($a_{\mathrm{G}} = 2.46 ~\AA$) is aligned to the hexagonal boron nitride ($a_{\mathrm{hBN}} = 2.5025~ \AA$), the lattice mismatch ($\epsilon = a_{\mathrm{G}}/a_{\mathrm{hBN}} -1\approx 1.7\%$) generates a moire pattern leading to secondary Dirac cone features~\cite{yankowitz2012, gbn1, gbn2}
and a primary Dirac point band gap forms\cite{amet, hunt} due to the moire pattern strains~\cite{woods2014, jung2015GBN, sanjose3}. 
Those moire strain profiles are tunable with twist angle $\theta = 0^{\circ} \sim 1^{\circ}$ leading to superlattice lengths ${\ell_m} \approx {a_G}/{\sqrt{\epsilon^2 + \theta^2}} \approx$ 14$\sim$10~nm that modifies band properties of the aligned single layer graphene and multilayer graphene~\cite{leconte2017,nanolettKIM,leconte2019commensurate,guorui1,guorui2,chittari,senthil,guorui2020,david2021,jixiang2022,bascones2022}. 
 The aligned ABC stacked rhombohedral trilayer graphene on hexagonal boron nitride (3LG/BN)~\cite{guorui1,guorui2,chittari,senthil,guorui2020,david2021,jixiang2022,bascones2022} is one of the important systems that develop interface-interaction induced band isolation where a perpendicular electric field can achieve narrow bandwidths comparable to the Coulomb energy ($U\approx 25~$meV)~\cite{guorui1,guorui2,guorui2020,jixiang2022}, making it a powerful platform for studying flat band driven phenomena.
In this manuscript we demonstrate that in rhombohedral four-layer graphene on boron nitride (4LG/BN) and $n$LG/BN systems up to $n=8$ we can achieve even narrower nearly flat bands than those of 3LG/BN whose bandwidths are typically $W\approx30\sim40~$meV in the absence of an electric field. 
While the low energy bandwidths become progressively narrower for increasing number of layers $n$ in rhombohedral multilayers, with 4LG narrowing between a factor two to four over 3LG, comparison against 5LG, 6LG, 7LG, and 8LG systems aligned on hexagonal boron nitride in the absence of perpendicular electric fields reveals that 5LG systems can already host optimally narrowest bands comparable to those of 6$\sim$8LG depending on the specific moire substrate potential used. 
The manuscript is structured as follows. 
In Sec.~II we present the full-bands continuum model Hamiltonian, In Sec.~III 
we discuss the results on the bandwidths, Chern numbers and local density of states as a function of twist angles and electric fields,
and in Sec.~IV we present the summary and conclusions. Extended data for $n=5\sim8$ systems are presented in the appendix. 

\section{Model Hamiltonian} 
\label{hamiltonian_model}
\begin{figure*}[!htb]
\includegraphics[width=17.5cm,angle=0]{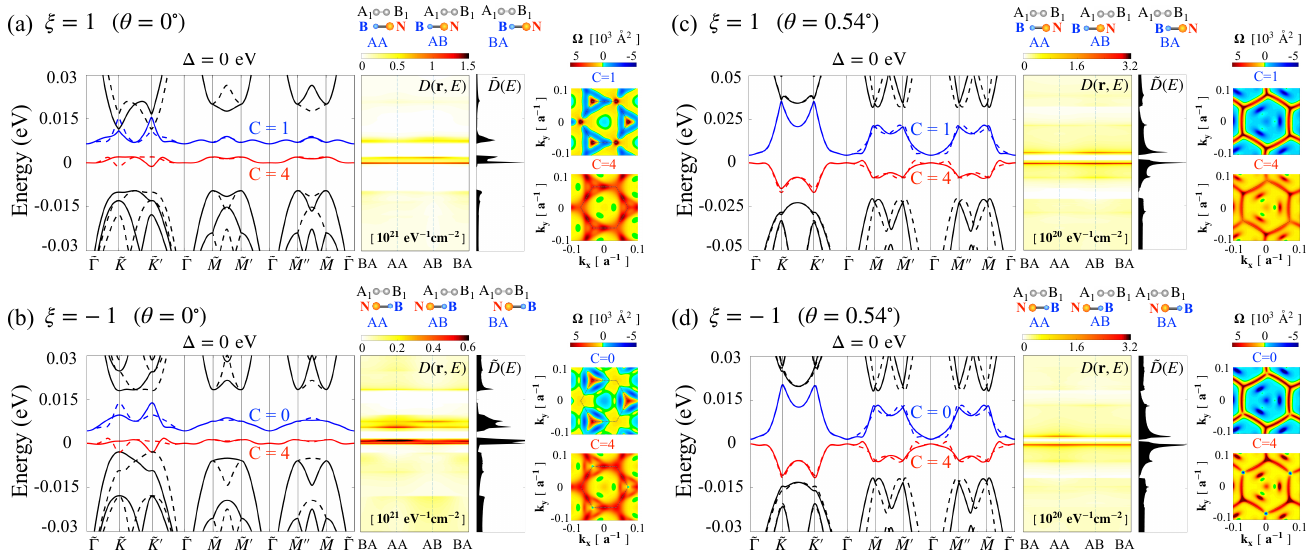} 
\caption{(color online) 
We present the band structure plots of 4LG/BN for two hBN orientations ($\xi = \pm 1$) at twist angles of $\theta = 0^{\circ}$ and $\theta = 0.54^{\circ}$, with $\Delta = 0$ eV and for valley $\nu = 1$ ($\nu = -1$) shown using solid (dashed) lines. To facilitate analysis, we use a compact representation of the local density of states (LDOS), $D({\bm r}, E)$, along the local commensurate stacking (AA, AB, and BA). We observe that the Van Hove singularities (vHS) are either localized to AB (AA) stacking for $\xi = 1$ ($\xi = -1$), or spread within the moiré unit cell. Delocalization occurs at the isolated valence bands, where the normalized density of states (DOS), $\tilde{D}(E) = D(E)/{\rm max}(D(E))$, clearly shows their band isolation. Additionally, we calculate the Berry curvatures ($\Omega$) of the valence (red) and conduction (blue) bands for valley $\nu = +1$. The unequal positive and negative weights of Berry curvatures near the MBZ corners result in non-zero valley Chern numbers, $C_{C1}^{+} = 1$ and $C_{V1}^{+} = 4$.
}
\label{Fig:Bands}
\end{figure*}

We model the rhombohedral tetralayer graphene on hexagonal boron nitride (4LG/BN) as
\begin{equation}
 \begin{aligned}
 {H} = {H}_{4LG} + {H}_M
 \end{aligned}
 \label{Eq:Hamiltonian}
 \end{equation}
as a sum of the four layers Hamiltonian $H_{4LG}$ and the effective intralayer moire pattern potential $H_M$ in the graphene layer contacting BN.
The $H_{4LG}$ term is a full-bands tight binding bands model 
\begin{equation}
 \begin{aligned}
 H_{4LG} = 
 \begin{pmatrix} H_{11} & H_{12} & H_{13}& {\bm 0}_{2 \times 2}\\ 
H_{12}^{\dagger}&H_{22} & H_{12}& H_{13} \\
H_{13}^{\dagger}&H_{12}^{\dagger}& H_{33} & H_{12} \\
{\bm 0}_{2 \times 2}&H_{13}^{\dagger}&H_{12}^{\dagger}& H_{44} \end{pmatrix}, 
 \end{aligned}
 \label{Eq:Ham4lg}
 \end{equation}
whose hopping parameters are chosen to match the LDA density functional theory
where the intra-$(H_{ll})_{2 \times 2}, (l=1,2,3,4)$ and inter-layer 
$(H_{ij})_{2 \times 2},(i \ne j=1,2,3)$ hamiltonian terms are given by
\begin{equation}
 \begin{aligned}
H_{ll}({\bm k}) &= \begin{pmatrix} u_{A_l} & \upsilon_0 \pi^{\dagger} \\ \upsilon_0 \pi & u_{B_l} \end{pmatrix} + V_{ll} \mathbb{1}, \\ 
H_{12}({\bm k}) &= \begin{pmatrix} - \upsilon_4 \pi^{\dagger} & -\upsilon_3 \pi \\ 
 t_1 & -\upsilon_4 \pi^{\dagger}\end{pmatrix}, \\
H_{13}({\bm k}) &= \begin{pmatrix} 0 & t_2 \\ 0 & 0 \end{pmatrix}.
 \end{aligned}
 \label{Eq:Hamiltonian2}
 \end{equation}
where $\pi = (\nu p_x + i p_y )$ is defined in terms of
the valley index $\nu = \pm1$ using the momentum vector ${\bm p} = (p_x, p_y)$ 
measured from the principal Dirac points ${\bm K}_{\nu} =\left( \nu \frac{4 \pi}{3 a_G}, 0\right)$.
The intralayer nearest neighbor sublattices ($A_l ~\& ~B_l$ where $l = 1, 2, 3, 4$ is the layer index) are 
connected through the $t_0 = -3.1$~eV hopping term whose magnitude is slightly larger than 
$t_0 \simeq -2.6~$eV obtained from density functional theory (DFT) local density approximation (LDA)
to partially account for the Coulomb interaction-driven Fermi velocity enhancement.
We use the Fermi velocity parameters $\upsilon_i = ( \sqrt{3} a/2 \hbar ) |t_i|$ in the Hamiltonian. 
The term $V_{ll} \mathbb{1}$ is used to introduce the interlayer potential difference ($\Delta$) between contiguous layers through a 
perpendicular external electric field. For convenience we use equal magnitude potential drops
proportional to $\Delta$ given by
\begin{equation}
V = \Delta \left( \frac{3}{2}, \frac{1}{2}, -\frac{1}{2}, -\frac{3}{2} \right)
\end{equation}
such that the interlayer potential difference between top and bottom graphene layers is given by $(n-1) \Delta$ where $n = 4$ for a tetralayer.
The remote hopping term between the interlayer sublattices $A_l ~\& ~B_{l+1}$ is
$t_3 = 0.293$~eV, and between $A_l(B_l)~\& ~A_{l+1}(B_{l+1})$ is given by $t_4 = 0.144$~eV, see Fig.~\ref{Fig:Bands}. 
The hopping energy between the adjacent interlayer vertical sublattices ($B_l ~\& ~A_{l+1}$) is 
$t_1 = 0.3561$~eV, and for the sublattices ($A_l ~\& ~B_{l+2}$) the hopping energy is $t_2 = -0.0083$~eV. 
The diagonal site potentials $u_{A_l}$ ($u_{B_l}$) at each sublattice of 4LG are
\begin{equation}
 \begin{aligned}
 u_{A1} &= u_{B4} = 0~\textrm{eV}, \\  
 u_{B1} &= u_{A4} = 0.0122~\textrm{eV}, \\
 u_{A2/A3} &= u_{B2/B3} = -0.0164~\textrm{eV}. \\
 \end{aligned}
 \end{equation}
In Fig.~\ref{Fig:DFT_compare} we confirm the close agreement of the DFT-LDA bands 
with the full-bands tight binding model from 3LG up to 8LG especially in the low energy range of 
$\pm 0.05~$eV and near the Dirac point $|{\bm k}|<0.05\left({4 \pi} / {3 a_G}\right)$. 
\begin{figure}[!htb]
\includegraphics[width=8.5cm,angle=0]{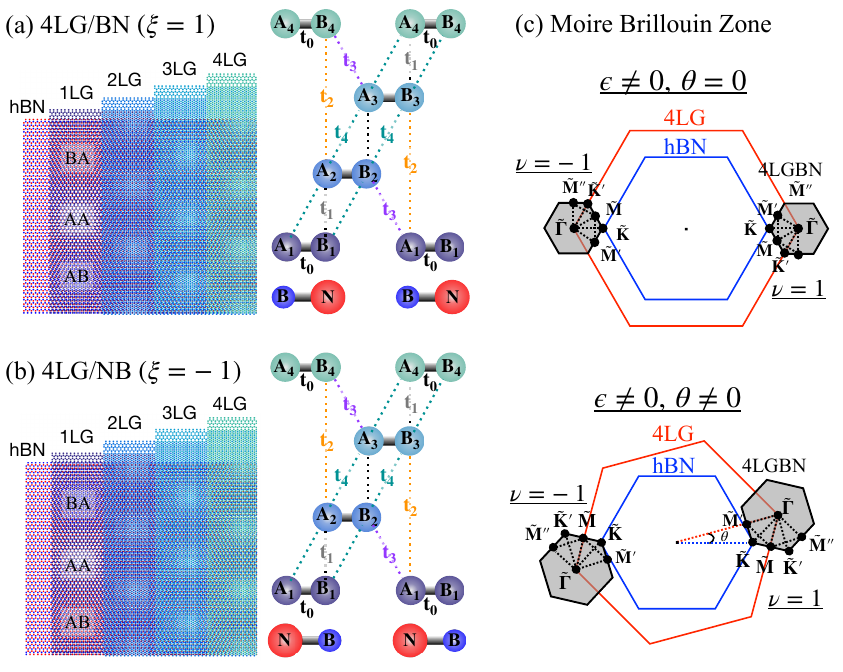} 
\caption{(color online) 
The atomic structure of four-layer graphene (4LG) on boron nitride moire superlattices with two possible hexagonal boron nitride (hBN) orientations is shown in (a) $\xi = 1$ (4LG/BN) and (b) $\xi = -1$ (4LG/NB). The moire superlattice formation due to lattice mismatch is illustrated for the graphene layers (1LG, 2LG, 3LG, and 4LG) where the local commensurate stacking are illustrated with AA, AB, BA labels. We show the local atomic structures illustrating the rhombohedral stacking sublattices A$_l$ and B$_{l}$, where $l = 1,2,3,4$ is the layer index, together with the intralayer ($t_0$) and interlayer ($t_1,~t_2,~t_3,~t_4$) nearest neighbor hopping terms. 
(c) We illustrate the moire Brillouin zone (MBZ) without (upper) and with (lower) a twist angle next to the Brillouin zones (BZ) of graphene (red hexagon) and hBN (blue hexagon) layers. The shaded hexagons indicate the MBZ represented at the two valleys $\nu = \pm 1$ of graphene. We define the Dirac points of the pristine 4LG as the center of the MBZ ($\tilde{\Gamma}$), and one of the corners of MBZ ($\tilde{K}$) corresponds to the Dirac points of hBN for each valley $\nu = \pm 1$.
}
\label{Fig:schematic}
\end{figure}
\par We capture the effect of G/BN moire superlattice by adding the effective intralayer 
moire potential $H^{\xi}_M$ acting at the bottom layer of 4LG 
\begin{equation}
 \begin{aligned}
H^{\xi}_{M}( {\bm r})
= &V^{\xi}_{AA}({\bm r}) \left(\frac{\mathbb{1}+\xi \sigma_z}{2}\right) + V^{\xi}_{BB}({\bm r}) \left(\frac{\mathbb{1}-\xi { \sigma}_z}{2}\right) \\&+ V_{BA}^{\xi}({\bm r})\cdot { \sigma}^{\xi}_{xy} \, \delta_{\nu,1} + V_{AB}^{\xi}({\bm r})\cdot { \sigma}^{\xi}_{xy} \, \delta_{\nu,-1}.
 \end{aligned}
 \label{Eq:moire}
 \end{equation}
Here, ${\bm \sigma}^{\xi}_{xy} = \left(\sigma_x, \xi \sigma_y\right)$ and $\sigma_z$ are the Pauli matrices in the sublattice basis. 
The $ 0^{\circ}$ and $ 60^{\circ}$ orientations of the hBN substrate give rise to different moire interlayer potentials. 
We use the $\xi=1$ label for the $0^{\circ}$ orientation of hBN sheet that for $AA$ stacking the B and N atoms are right below $A_1$ and $B_1$ sites respectively,
while the AA stacking for $\xi = -1$ corresponding to $ 60^{\circ}$ orientation 
and the N and B atoms are below $A_1$ and $B_1$ sites, see Fig.~\ref{Fig:schematic}(a) $\&$ (b).
\par The modifications of the onsite potential and inter sublattice interaction in the bottom layer graphene of 4LG contacting hBN to capture the effective intralayer moire potential is given by the following equations:
\begin{equation}
V^{\xi}_{AA(BB)}({\bm r}) = 2\, C_{AA(BB)} \mathrm{Re}\left[ e^{i \phi_{AA(BB)}} \, f^{\xi}({\bm r})\right],
 \end{equation}
\begin{equation}
 \begin{aligned} 
V^{\xi}_{BA}({\bm r}) &=\left(V^{\xi}_{AB}({\bm r})\right)^{*}
= 2\, C_{AB}\left( \mathrm{cos}(\phi) \hat{z}\times\mathbb{1} - \mathrm{sin}(\phi) \mathbb{1}\right)  \\ 
&\times \left( \frac{1}{| {\tilde{\bm G}_1}|} \nabla \mathrm{Re} \left[ e^{i (-\phi_{AB}+\pi/6)} \, f^{\xi}({\bm r})\right]\right),
\end{aligned}
\end{equation}
where the coefficients adopted from Ref.~\cite{jung2014ab} are $C_{AA} = -14.88 $~meV, $\phi_{AA} = 50.19 ^{\circ}$, $C_{BB} = 12.09$~meV, $\phi_{BB} = -46.64 ^{\circ}$, $C_{AB} = 11.34 ~$~meV, $\phi_{AB} = 19.60 ^{\circ}$, 
and the function as $f^{\xi}({\bm r})=\sum_{m=1}^{6}{ \left( \frac{1+(-1)^{m+1}}{2}\right) e^{-i \xi { \tilde{\bm G}_m} \cdot {\bm r}}}$. 
The moire reciprocal lattice vectors are defined as ${ \tilde{\bm G}_m}=\hat{R}_{\frac{\pi}{6}(m-1)} {\tilde{\bm G}_1}$ where ${ \tilde{\bm G}_1}=(\frac{a_{\mathrm{G}}}{a_{\mathrm{hBN}}} \hat{R}_{\theta}-\mathbb{1})^{\dagger}[0,\frac{4 \pi}{\sqrt{3}a_{\mathrm{G}}}] \approx (\epsilon \mathbb{1}-\theta \hat{z}\times\mathbb{1})[0,\frac{4 \pi}{\sqrt{3}a_{{\bm G}}}] $ for the lattice mismatch $\epsilon \approx -1.7\%$ and twist angle $\theta$ between the graphene and hBN ~\cite{jung2015GBN}. 
In Fig.~\ref{Fig:schematic}(c), we illustrated the moire Brillouin zone (MBZ) with the original Brillouin zone of the 4LG and hBN layers to show the effect of the lattice mismatch and twist angle. 
The reduced size of MBZ leads to a repulsion gap at the MBZ zone corners in the 4LG/BN and isolates the bands near the Fermi level.
\begin{figure*}[!htb]
\includegraphics[width=17.5cm,angle=0]{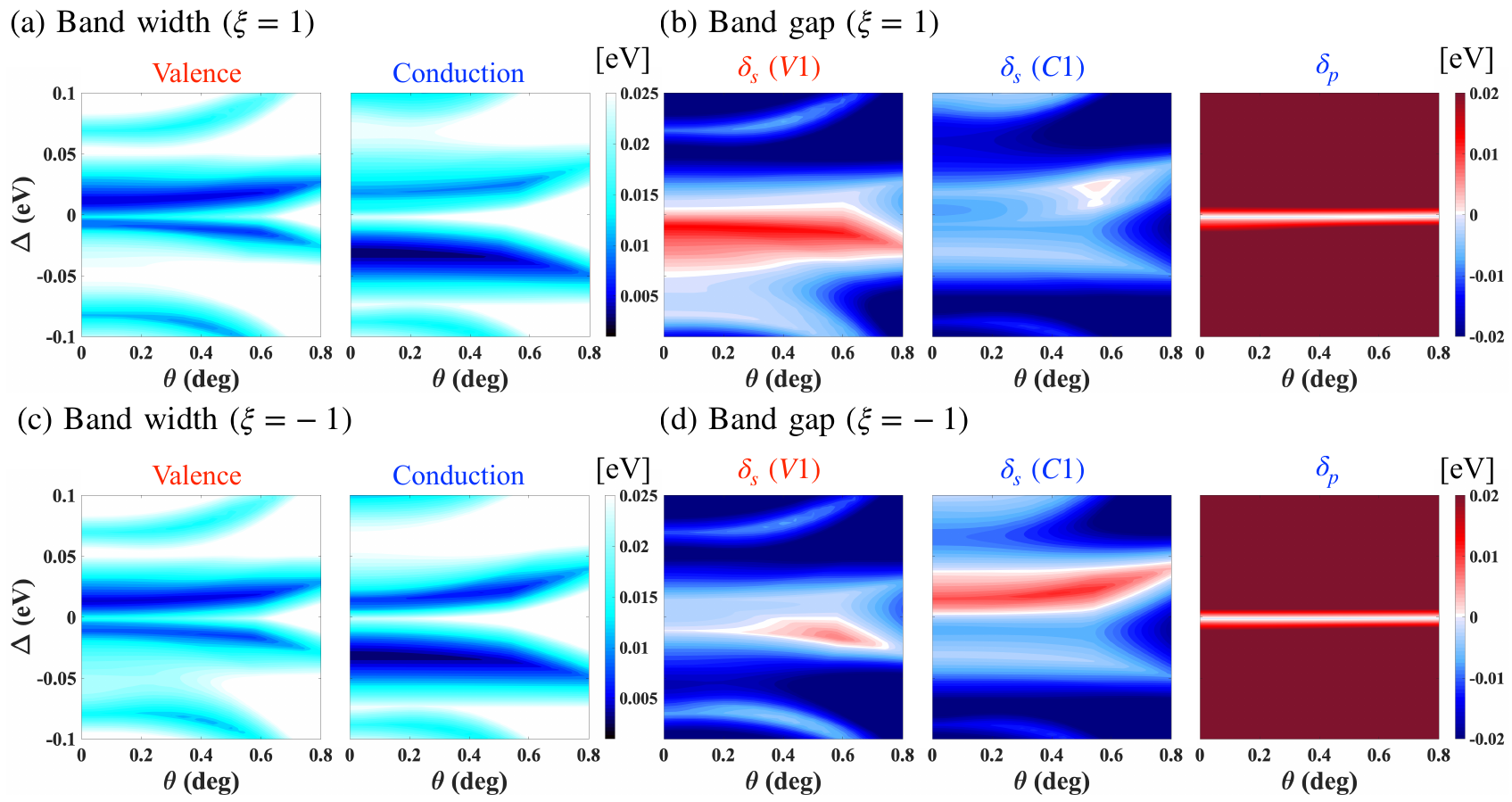} 
\caption{(color online) 
The low energy bandwidth variation for the parameter set of interlayer potential difference ($\Delta$) and twist angle ($\theta$) is presented for the two orientations of hBN, (a) $\xi =1$ and (c) $\xi = -1$. In (b) and (d), we present the corresponding primary ($\delta_p$) and secondary gaps for valence ($\delta_s~(V1)$) and conduction ($\delta_s~(C1)$) bands, to estimate the isolation of valence and conduction bands at CNP.
}
\label{Fig:BWGAP}
\end{figure*}
\section{Results and Discussion}
In this section we present the valley Chern numbers, and local density of states that are associated with the large $U_{\rm eff}/W$ values where strong correlations can be expected in the parameter space of twist angles and interlayer potential differences.

The low energy band structure of 4LG/BN have extremely narrow bands near the charge neutrality point (CNP) 
thanks to the $\sim p^{N}$ behavior of the band dispersion near the Dirac point 
of the minimal model rhombohedral multilayer graphene as the layer number $N$ increases~\cite{hmin}. 
More realistic band structures include distortions due to trigonal warping and electron-hole symmetry breaking
terms included in the remote hopping terms of the Hamiltonian as illustrated in the LDA bands 
from 3LG to 8LG in Fig.~\ref{Fig:DFT_compare} in the appendix. 
Addition of a moire pattern introduces a band gap both at the charge neutrality 
and near the MBZ corners that isolates the low energy valence and conduction bands from the adjacent 
bands even in the absence of an external electric field, see Fig.~\ref{Fig:Bands}. 
The band structures are shown for 4LG/BN for two different orientations of hBN,
namely 4LG/BN ($\xi = 1$) and 4LG/NB ($\xi = -1$), and we considered the aligned $\theta = 0^{\circ}$ and a small finite twist angle of $\theta = 0.54^{\circ}$. 
The low energy bands have bandwidths between 
$W\approx 12\sim20~$meV where the bandwidth of the first valence ($V1$) or conduction ($C1$) band is calculated as $W_{V1(C1)} =  E^{V1(C1)}_{\max}-E^{V1(C1)}_{\min} $. 
We note that the low energy valence and conduction bandwidths of 4LG/BN and all the way up to 8LG/BN
are smaller than those of 3LG/BN~\cite{chittari} (See Fig.~\ref{Fig:BS_5layer}(c) $\&$ (d)). 
\par 
The real-space distribution of the wave functions associated with the nearly flat bands in 4LG/BN 
gives an idea about the kind of broken symmetries that the system can host. 
We calculated the density of states (DOS) $D(E)$ through
\begin{equation}
 \begin{aligned}
D(E) 
&=\int_{\rm MBZ} \frac{d{\bm k}}{(2 \pi)^2}\,\sum_{n^{\prime}} |\psi_{n^{\prime}}({\bm k})|^2\, \delta(E - E_{n^{\prime}}({\bm k}))
 \end{aligned}
 \label{Eq:DOS}
 \end{equation}
and the local density of states (LDOS) $D({\bm r},E)$ through
 \begin{equation}
 \begin{aligned}
D({\bm r},E)
&= \int_{\rm MBZ} \frac{d{\bm k}}{(2 \pi)^2} \, \sum_{n^{\prime}}|\psi_{n^{\prime}}({\bm k} : {\bm r})|^2\, \delta(E - E_{n^{\prime}}({\bm k}))
 \end{aligned}
 \label{Eq:LDOS}
 \end{equation}
using information from the real-space wave functions 
$\psi_{n^{\prime}}({\bm k} : {\bm r}) = \sum_{\tilde{G}} \, \psi_{n^{\prime}}({ {\bm k} + \tilde{\bm G}}) \, e^{-i ({\bm k + \tilde{\bm G}})\cdot {\bm r}}$ 
where the band indices $n^{\prime}= Vn (Cn)$ refer to the $n^{\rm th}$ valence (conduction) bands counting from the charge neutrality point. In Fig.~\ref{Fig:Bands}, we show the compact representation of the LDOS, $D({\bm r},E)$ at each local commensurate stacking (AA, AB and BA), along with the normalized density of states defined as $\tilde{D}(E) = D(E)/{\rm max}(D(E))$. The density of states $\tilde{D}(E)$ plots show the van Hove singularities (vHS) of DOS peaks for the isolated low energy valence and conduction bands. Additionally, the LDOS plots show that the vHS of the conduction band is strongly localized at AB(AA) stacking for 
$\xi = 1$ ($\xi = -1$) at both $\theta = 0^{\circ}, 0.54^{\circ}$. 
However, the valence band states at the vHS distribute the carrier densities almost equally at all local commensurate stacking sites.
These differences in the localization behavior between the conduction and valence band states should in turn lead to very different spin or charge textures of the associated ordered phases upon inclusion of Coulomb interactions.

\par Furthermore, the low energy isolated bands show a non-trivial topological nature. We calculated the valley resolved Chern number ($C^{\nu}_{V1(C1)}$) of the first valence (conduction) band using the following equations:  
 \begin{equation}
 \begin{aligned}
 C_{n^{\prime}} &= \int_{\rm MBZ} d^2 {{\bm k}} \quad \Omega_{n^{\prime}}( { \bm {k}} ) /  (2 \pi),
 \end{aligned}
 \label{Eq:Chern}
 \end{equation}
where the Berry curvature $\Omega_{n^{\prime}}({\bm k})$~\cite{berry_rmp} is defined as
 \begin{equation}
 \begin{aligned}
 \Omega_{n^{\prime}}(\bm{{k}}) &= -2 \sum_{i \neq n^{\prime}} \mathrm{Im} 
 \left[\frac{\mel{u_{n^{\prime}}}{\frac{\partial H}{\partial {k}_x}}{u_{i}} \mel{u_{i}}{\frac{\partial H}{\partial {k}_y}}{u_{n^{\prime}}}}{(E_{i} - E_{n^{\prime}})^2}\right],
 \end{aligned}
 \label{Eq:Berry}
 \end{equation}
for the $n^{\prime} = Vn (Cn)$ valence (conduction) bands from the charge neutrality point
where $\ket{u_{n^{\prime}}}$ are the moire Bloch states of index $n^{\prime}$, and $E_{n^{\prime}}$ are the band energies. 
The calculated Berry curvatures for the low energy bands are represented in Fig.~\ref{Fig:Bands} for $\xi = \pm1$ and $\theta = 0^{\circ}, 0.54^{\circ}$. The Berry curvature plots have hot spots near the MBZ corners, and the unequal weights of Berry curvatures give rise to non-zero valley Chern numbers $C^{\nu=\pm 1}_{V1} = \pm ~4$ for the
low energy valence bands of 4LG/BN at twist angles $\theta = 0^{\circ}, 0.54^{\circ}$ with $\xi = \pm 1$, 
and we get $C^{\nu=\pm1}_{C1} = \pm~1$ for the low energy conduction bands for $\xi = 1$ but zero for $\xi = -1$.
We will discuss later how the valley Chern number phase diagram can change as a function of interlayer potential difference.

\subsection{Low energy bands}
\label{flatbands}

\par In the following, we discuss the bandwidth of the low energy bands near the Fermi level of 4LG aligned to hBN in the parameter space of twist angles ($\theta$) and interlayer potential differences ($\Delta$).
In our previous reports~\cite{srivani2019,tbbg,tMBG,chittari}, we systematically demonstrated that the interlayer potential difference introduced through a perpendicularly applied electric field is an effective control knob to tune the bandwidth and band isolation of the low energy bands. 
The interlayer potential difference ($\Delta$) results from unequal intralayer potentials $V_{ii} \mathbb{1}$ in each layer of 4LG/BN in Eq.~(\ref{Eq:Hamiltonian2}).
In Fig.~\ref{Fig:BWGAP}(a) and (c), we show the variation of bandwidth as a function of $\Delta$ and $\theta$. The on-site Coulomb repulsion energy for graphene on hBN is estimated to be $U\approx 25~$meV for the moire length of $\sim14.5~$nm. From the bandwidth phase diagram, it is evident that a large set of the parameter space of $\Delta$ and $\theta$ shows a bandwidth smaller than $\sim$25~meV. 
Further, we also investigate the isolation of the low energy flat bands from the adjacent remote bands by calculating the gaps at CNP ($\delta_p$) and secondary gaps ($\delta_s$) as a function of $\Delta$ and $\theta$ (see Fig.~\ref{Fig:BWGAP}). Here, the primary gap at CNP is defined as $\delta_p = E^{C1}_{\min} -E^{V1}_{\max}$ and the secondary gap (a gap between the valence (conduction) band from its higher energy bands) is defined as $\delta_s (n^{\prime}) = E^{n^{\prime}}_{\min} - E^{n^{\prime}}_{\max} $ with $n^{\prime} = V1 (C1)$. 
A positive value denotes the magnitude of the gap, and a negative value indicates the degree of band overlap. 
The low energy bands isolate when both secondary and primary gaps are simultaneously positive. 
The primary gap at CNP is always open practically for all values of $\theta$~and~$\Delta$ except for a specific value of $\Delta$,
see Fig.~\ref{Fig:BWGAP}. 

 A few important observations from the bandwidth and bandgap phase diagrams are summarized below. 
 (i) Very narrow valence (conduction) bands ($W<$ 5~meV) are possible for positive(negative) values of $\Delta \approx 0.012 \sim 0.013~{\rm eV}$ ($\Delta \approx -0.025\sim-0.038~{\rm eV}$) at zero-twist ($\theta = 0^{\circ}$) for both $\xi = \pm1$.
 (ii) The small twist angles $\theta = 0\sim0.8^{\circ}$ retain the smallest bandwidths ($W<$ 5~meV) for the valence and conduction bands. 
(iii) For $\xi = 1$, the conduction bands are isolated for the specific conditions of $\theta \approx 0.5\sim 0.6^{\circ}$ and $\Delta = 0.007 \sim 0.027$~eV while the valence bands are isolated for the considered range of $\theta=0\sim0.8^{\circ}$ for $\Delta =-0.05$ to $0.01~{\rm eV}$.
(iv) For $\xi =-1$, the conduction bands are isolated for the range of $\Delta=$0 to 0.04~eV and $\theta=0^{\circ} \sim 0.5^{\circ}$ while the valence bands are isolated when $\theta\approx 0.5 \sim 0.6^{\circ}$ and $\Delta=-0.004$ to $-0.02$~eV.
From the bandwidth and band gap phase diagrams we can observe that the narrowest bandwidths of $W=$ 5$\sim$10~meV for the isolated valence (conduction) bands is associated with $\Delta = -0.009~(0.009)$ eV for $\xi =1$ and $\Delta = -0.011~(0.015)$~eV for $\xi =-1$ for the twist angle range $\theta = 0^{\circ} \sim 0.5^{\circ}$. See Fig.~\ref{Fig:BS_0deg} where we summarize band structures for these selected system parameters. These observations help us draw a few important conclusions on the possible Coulomb interation driven phases of 4LG/BN that we will discuss further in the following subsection.

\subsection{Nearly flat bands ($U_{\rm eff}/W\gg1$ )}
\label{strongU}

\begin{figure}[!htb]
\includegraphics[width=7.5cm,angle=0]{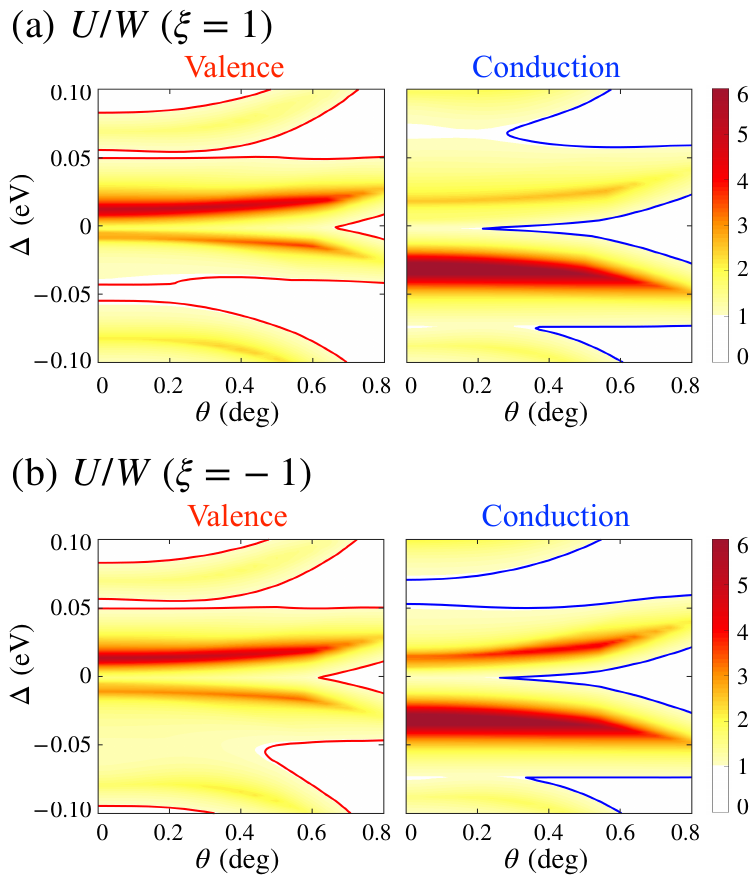} 
\caption{(color online) 
We present a phase diagram showing the ratio between the bare on-site Coulomb repulsion energy ($U$) and low energy band bandwidths ($W$) as a function of an interlayer potential difference ($\Delta$) and twist angles ($\theta$) for two orientations of hBN: (a) $\xi =1$ and (b) $\xi = -1$. The colored region represents the narrow bands where $U/W>1$, indicating strong electronic correlations. We find that almost all the system parameters of twist angle $\theta=0\sim0.8^{\circ}$ and interlayer potential difference $|\Delta|<0.1~$eV exhibit narrow bands satisfying $U/W>1$ for both $\xi=\pm1$ hBN orientation, suggesting that strong electronic correlations are present in a wide range of experimental conditions.
}
\label{Fig:UeffU_W}
\end{figure}

Here we compare the on-site Coulomb repulsion energy ($U$) and bandwidth ($W$) of valence and conduction bands in search for the narrow bandwidth regime with strong effective Coulomb interactions where $U_{\rm eff}/W\gg1$. The greater the $U$ compared to the bandwidth $W$ ($U/W>1$) indicates higher chances of finding Coulomb interaction induced correlated phases. 
We define the on-site Coulomb repulsion energy $U$ using the moire superlattice constant ${\ell_m} \approx a_G/\sqrt{\epsilon^2 + \theta^2}$ and a relative permittivity $\epsilon_{r}=4$
\begin{equation}
 \begin{aligned}
 U &= \frac{e^{2}}{4 \pi \epsilon_{r} \epsilon_{0} {\ell_m}},
 \end{aligned}
 \label{Eq:U}
 \end{equation}
where $e$ is the electron charge, and $\epsilon_0$ is the permittivity of vacuum. In Fig.~\ref{Fig:UeffU_W}, we summarized the ratio ($U/W$) of the low energy bandwidth $W$ versus the on-site Coulomb repulsion energy $U$ as a function of $\theta$ and $\Delta$, where the colored area manifests the region for the possible narrow bands ($U/W >1$).
Compared to the estimated Coulomb repulsion energy($\approx 25$~meV), the bandwidths of valence and conduction bands in 4LG/BN are small enough to satisfy the $U/W > 1$ condition for most of the considered parameter space.
\begin{figure}[!htb] 
\includegraphics[width=8.5cm,angle=0]{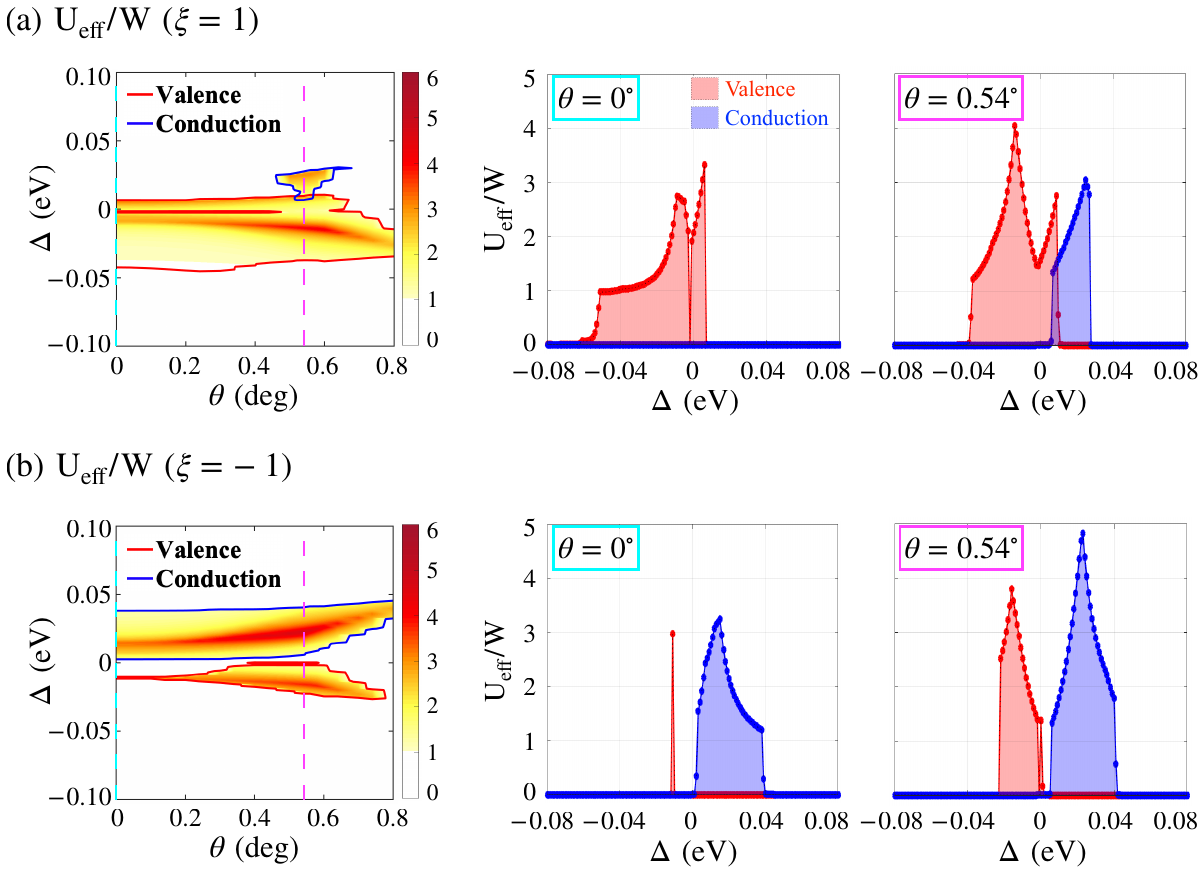} 
\caption{(color online) 
We present a phase diagram illustrating the variation of the ratio between effective on-site Coulomb repulsion energy ($U_{\rm eff}$) and bandwidths ($W$) as a function of interlayer potential difference ($\Delta$) and twist angles ($\theta$) for two orientations of hBN: (a) $\xi = 1$ and (b) $\xi = -1$. On the right side of the diagram, we show $U_{\rm eff}/W$ results at two specific twist angles: $\theta = 0^{\circ}$ (cyan) and $\theta = 0.54^{\circ}$ (magenta).
For $\theta=0^{\circ}$, $U_{\rm eff}/W > 1$ is only achievable for the valence band with negative $\Delta$ for $\xi=1$, and for the conduction band with positive $\Delta$ for $\xi=-1$. However, at $\theta = 0.54^{\circ}$, both valence and conduction bands exhibit $U_{\rm eff}/W\gg1$, with negative $\Delta$ required for the valence band and positive $\Delta$ for the conduction band.
Our results reveal that $U_{\rm eff}/W>1$ for both conduction and valence bands can be achieved at $\Delta=0.009~$eV for $\xi=1$ and $\theta=0.54^{\circ}$. 
}
\label{Fig:Ueff_05}
\end{figure}

We further consider the screening effects by adding an exponential decay term $e^{ -{\ell_m} / \lambda_{D}}$~\cite{tbbg} adequately when there is band overlap. Here, the screened effective on-site Coulomb repulsion energy ($U_{\rm eff}$) is redefined as:
\begin{equation}
 \begin{aligned}
 U_{\rm{eff}} &= U~\textrm{exp}(- {\ell_m} / \lambda_{D}),
 \end{aligned}
\label{Eq:Ueff}
 \end{equation}
where $\lambda_{D} = 2 \epsilon_{0} /e^{2} D(\delta_{p}, \delta_{s})$ is the Debye length, which includes the 2D density of states $D(\delta_{p}, \delta_{s}) = 4 [|\delta_{p}|\, u(-\delta_{p}) \,+\, |\delta_{s}|\, u(-\delta_{s})]/(W^{2} A_{M})$. 
$A_{M} = \sqrt{3}{\ell_m}^2 / 2$ is the moire unit cell area, $W$ is the bandwidth,
and $u(x)$ is the Heaviside step function such that $u(-\delta_{p(s)})$) enhances the screening 
in the presence of band overlap ($\delta_{p(s)}<0$). 
We show that the screening effect due to the band overlap reduces the narrow bandwidth regime ($U_{\rm eff}/W>1$), see Fig.~\ref{Fig:Ueff_05} where we summarized $U_{\rm eff}/W$ for the parameter space of interlayer potential difference $\Delta$ and twist angle $\theta$, and specifically for $\theta = 0^{\circ}$ (cyan) and $~0.54^{\circ}$ (magenta). 
For $\xi=1$ (Fig.~\ref{Fig:Ueff_05} (a)), the valence band show $U_{\rm eff}/W \gg1$  for a large span of the parameter spaces ($\Delta\approx -50\sim10$~meV and $\theta=0\sim0.8^{\circ}$) while for the conduction band $U_{\rm eff}/W>1$ is only possible for the small island at $\theta\approx0.5\sim0.6^{\circ}$~and ~$\Delta\approx20\sim25$~meV.
In another substrate configuration, $\xi = -1$ (Fig.~\ref{Fig:Ueff_05} (b)), the conduction band has greater chances of achieving the flat bands regime at small positive $\Delta=0\sim40$~meV compared to the valence band, which has a limited area around $\Delta\approx 20$~meV~and~$\theta=0.4\sim0.6^{\circ}$.

Moreover, as shown in the phase diagram, the 4LG/BN flat bands have different particle-hole asymmetric behavior for different $\theta$. We found that at $\theta=0^{\circ}$ the valence (conduction) band only satisfies $U_{\rm eff}/W \gg1$ for $\xi =1$($\xi=-1$) substrate orientation. On the contrary, at $\theta=0.54^{\circ}$, both valence and conduction flat bands are possible, which requires
negative $\Delta$ for valence bands and positive $\Delta$ for the conduction bands.
Also, the maximum values of $U_{\rm eff}/W$ are observed when $\theta=0.54^{\circ}$ giving $U_{\rm eff}/W = 3.98$ for the valence band with $\xi=1$ orientation and $U_{\rm eff}/W = 4.7$ for the conduction band with $\xi =-1$. Hence, we suggest that $\theta = 0.54^{\circ}$ for the 4LG/BN device has greater chances of achieving Coulomb-interaction driven ordered states in a wider range of interlayer potential differences.

\begin{figure}[!htb]
\includegraphics[width=8.5cm,angle=0]{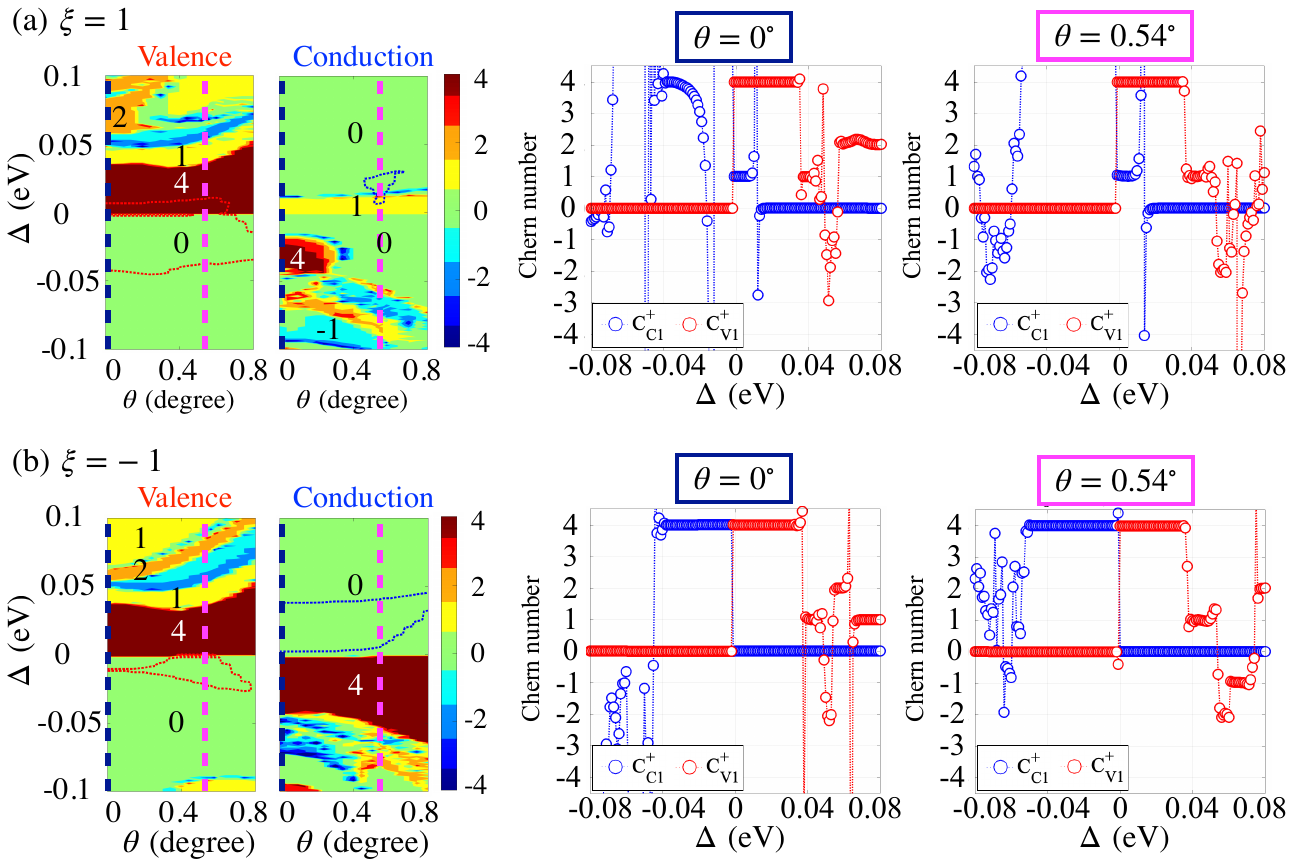} 
\caption{(color online) 
The valley Chern number of the low-energy valence and conduction bands is plotted as a function of interlayer potential difference ($\Delta$) and twist angle ($\theta$) for two substrate orientations: (a) $\xi=1$ and (b) $\xi=-1$. The dashed lines indicate two specific twist angles, $\theta=0^{\circ}$ (navy) and $\theta=0.54^{\circ}$ (magenta), and the phase transition between different Chern numbers is shown on the right-hand side of the phase diagram. The red (valence) and blue (conduction) dashed lines enclose regions where the effective on-site Coulomb repulsion energy $U_{\rm eff}$ is greater than the band widths $W$, i.e., $U_{\rm eff}/W>1$.
}
\label{Fig:Chern}
\end{figure}

\subsection{Valley Chern numbers}
\label{valleychern}
When the moire bands are isolated through gaps, they can acquire finite valley resolved Chern numbers~\cite{chittari,senthil}. 
The lowest energy bands of 4LG/BN in the absence of an external electric field have finite valley Chern numbers for both valence and conduction bands when $\xi = 1$, and for valence bands only for $\xi = -1$, or in other words the valence band 
has a finite valley Chern number $C_{V1}=4$ for both $\xi=\pm1$ substrate orientations, while 
the conduction band has the valley Chern number $C_{C1}=1$ only for $\xi = 1$. 

Here we analyze the phase diagram of the valley Chern numbers in the parameter space of interlayer potential differences $\Delta$ and twist angles $\theta$ for the two different types of hBN substrate orientations labeled through $\xi = \pm1$, specifically for $\theta = 0^{\circ}$ (navy dashed) and $~0.54^{\circ}$ (magenta dashed) that we summarize in Fig.~\ref{Fig:Chern}.
Upon application of a perpendicular electric field, the valence band valley Chern number $C_{V1}=4$ for $\xi = 1$ (Fig.~\ref{Fig:Chern} (a)) remains within a large set of parameter spaces of $\Delta \approx 0\sim35$~meV and $\theta=0\sim0.8^{\circ}$ and for the larger electric field switches to $C_{V1}=\pm1, \pm2$ at zero twist ($\theta=0^{\circ}$) and $C_{V1}=\pm1, 2$ at $\theta=0.54^{\circ}$.
The conduction band, however, shows a nonzero valley Chern number $C_{C1}=4$ only for the small angles $\theta\approx 0\sim0.2^{\circ}$ with negative $\Delta\approx-20\sim-40$~meV and $C_{C1}=1$ for positive $\Delta\approx 0\sim10$~meV.
In another substrate orientation, $\xi = -1$, (Fig.~\ref{Fig:Chern} (b)), the valence band shows $C_{V1}=4$ with $\Delta=0\sim35$~meV and moves to $C_{V1}=\pm1, \pm2$ with a larger positive interlayer potential difference, while the nonzero conduction band valley Chern number $C_{C1}=4$ only appears for the negative $\Delta=-40\sim0$~meV.
Hence, a single 4LG/BN device can have various topological states ($C_{V1}= \pm1,\pm2,4$, $C_{C1}=1, 4$), easily controlled through small changes in interlayer potential differences of the order of $|\Delta|<100 $~meV. 

Interestingly, we also found that rhombohedral stacked $n$LG/BN ($n=3\sim8$ layer graphene) for both substrate orientations ($\xi = \pm1$) have a finite valence band valley Chern number, $C_{V1} = n$ when $\Delta=0$~eV, see Fig.~\ref{Fig:BS_5layer}.
Similar to 4LG/BN, the conduction band of $\xi = -1$ have zero valley Chern number, $C_{C1}=0$ while multilayer 3LG/BN$\sim$8LG/BN systems for $\xi=1$ show various valley Chern numbers, $C_{C1}=2, 1, 1, -2,$ 0, and 0 at $\Delta = 0$~eV. 
Our calculations suggest that rhombohedral $n$LG/BN multilayers will be an excellent platform to explore a variety of topological phases with flat enough bands to find strong correlation effects.

\begin{figure*}[!htb]
\includegraphics[width=16.5cm,angle=0]{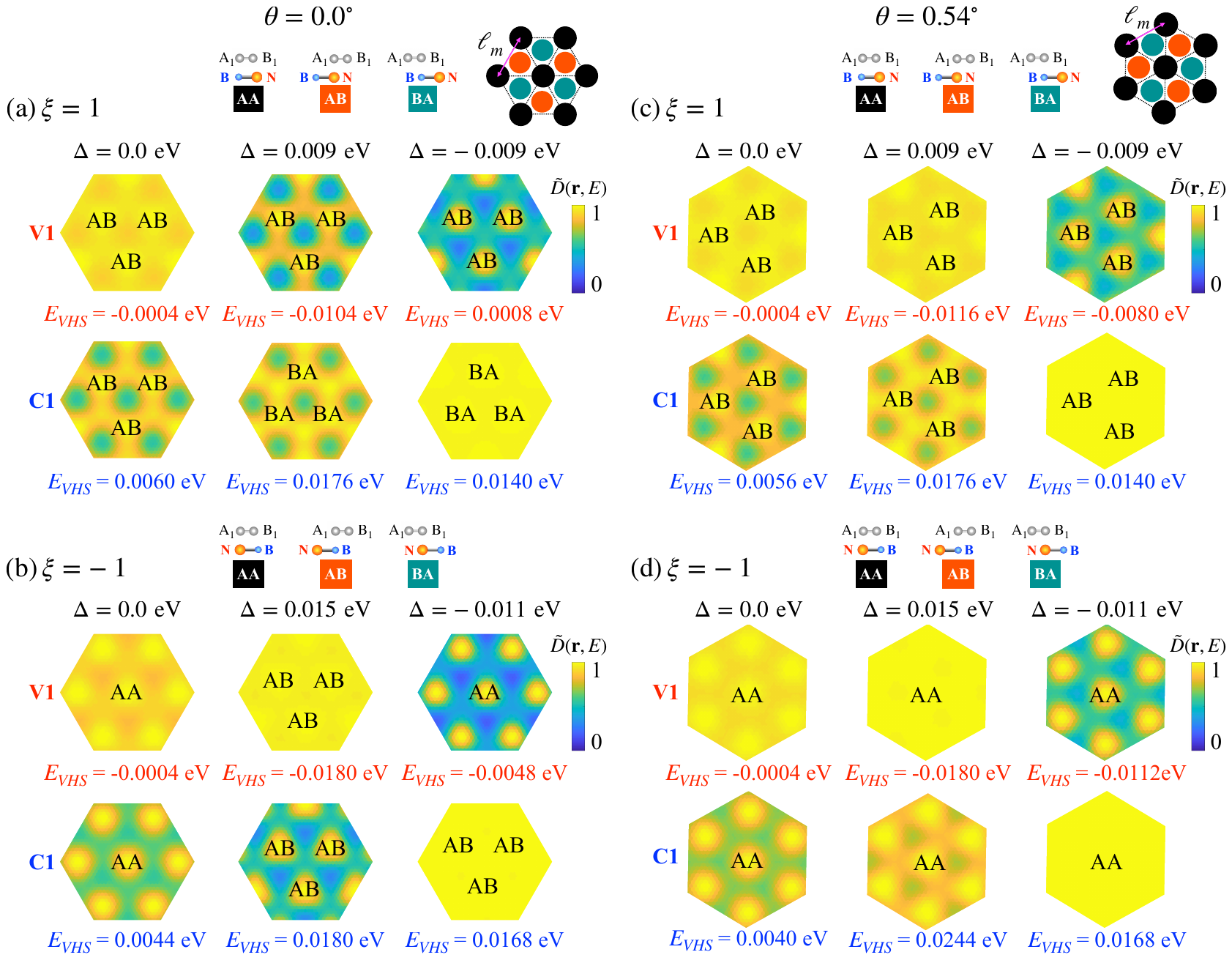} 
\caption{(color online) 
We present the normalized local density of states (LDOS) of the isolated valence ($V1$) and conduction ($C1$) bands of 4LG/BN at their van Hove singularities (vHS), which are defined as $\tilde{D}({\bm r},E) = D({\bm r},E)/{\rm max}(D({\bm r},E))$. The LDOS is plotted for two twist angles, $\theta = 0^{\circ}$ (left) and $\theta = 0.54^{\circ}$ (right), and two orientations of the hBN substrate, (a)-(b) $\xi =1$ and (c)-(d) $\xi =-1$.
We show the normalized LDOS for selected interlayer potential differences $\Delta = 0.0, 0.009, -0.009~\rm eV$ for $\xi =1$, and $\Delta = 0.0, 0.015, -0.011~\rm eV$ for $\xi = -1$. The energy value at the vHS ($E_{vHS}$) is listed at the bottom of each LDOS panel, and we label the LDOS maxima as AA, AB, or BA according to the local stacking regions defined at the top right corner of panels (a) and (c).
Our results show that the LDOS distributions at the vHS are influenced by system parameters such as the substrate orientation $\xi$, the interlayer potential difference $\Delta$, and the twist angle $\theta$, and exhibit distinct features for the valence and conduction bands.
}
\label{Fig:LDOS}
\end{figure*}

\subsection{Local density of states}
\label{ldos}
In this subsection, we discuss the local density of states (LDOS) of 4LG/BN at the van Hove singularities (vHS) of  
the low energy valence and conduction bands that are most prone to form ordered phases. 
Knowledge of their localization centers is helpful for anticipating and understanding the nature of the ground states that we can expect when Coulomb interactions are accounted for. 
We show that the LDOS profiles associated to the valence or conduction bands can be tuned by varying the system parameters such as electric fields, the twist angles, and the orientation of the hBN substrate. 
Our results are summarized in Fig.~\ref{Fig:LDOS}, where we show the unit renormalized LDOS function $\tilde{D}({\bm r},E)$ for twist angles $\theta = 0^{\circ},\,0.54^{\circ}$ for both orientations of the hBN substrate $\xi = \pm 1$. 
We explicitly label with AA, AB, and BA the local stacking configurations where the LDOS profiles show maxima values. 
For zero electric fields and $\xi = 1$ substrate orientation we notice that the LDOS accumulates at the AB local stacking sites, while for $\xi = -1$ the localization center switches to AA stacking sites for both valence and conduction bands. 
This behavior is observed for both $\theta = 0^{\circ}, \, 0.54^{\circ}$ twist angles considered and can be modified by applying an interlayer potential difference $\Delta$.

In the following, we discuss in more detail how an interlayer potential difference can alter the localization properties. 
The electron accumulation at the topmost layer favored by $\Delta > 0$ should predominantly 
concentrate at the low energy $B_4$ site, and likewise for $\Delta < 0$ will concentrate mostly at $A_1$, and in opposite senses for holes. 
We will show that $\Delta$ can modify the degree of localization of the states around particular local stacking configurations. 
Let us consider the $\xi = 1$ orientation for the hBN substrate (Fig.~\ref{Fig:LDOS}(a)) and focus on the $\theta = 0^{\circ}$ case for sake of definiteness. 
The valence band LDOS spread over all stacking configurations and are delocalized, mildly concentrating at the AB regions when $\Delta = 0$~eV. 
The initially widespread LDOS profile concentrates at the AB sites in 
the presence of interlayer potential differences either positive or negative, see the plots for $\Delta =  0.009$, $-0.009$~eV. 
We thus expect that a finite $\Delta$ will enhance the chances of triggering a Coulomb driven transition. 
For the conduction band, the LDOS concentrate initially at both AB to BA regions when $\Delta \neq 0$, where AB is slightly favored. 
A positive $\Delta = $0.009~eV shifts the carrier densities from AB to predominantly concentrate at the BA regions and also increases
the population of the AA regions, while for negative $\Delta = -0.009$~eV the LDOS profile spreads almost uniformly for all possible stacking regions. 
Hence, vertical fields allow transitions from a delocalized to AB concentrated LDOS profiles in the valence bands, 
or from AB/BA centered profiles to delocalized phases for conduction bands depending on applied electric fields.

\begin{figure}[!htb]
\includegraphics[width=8.5cm,angle=0]{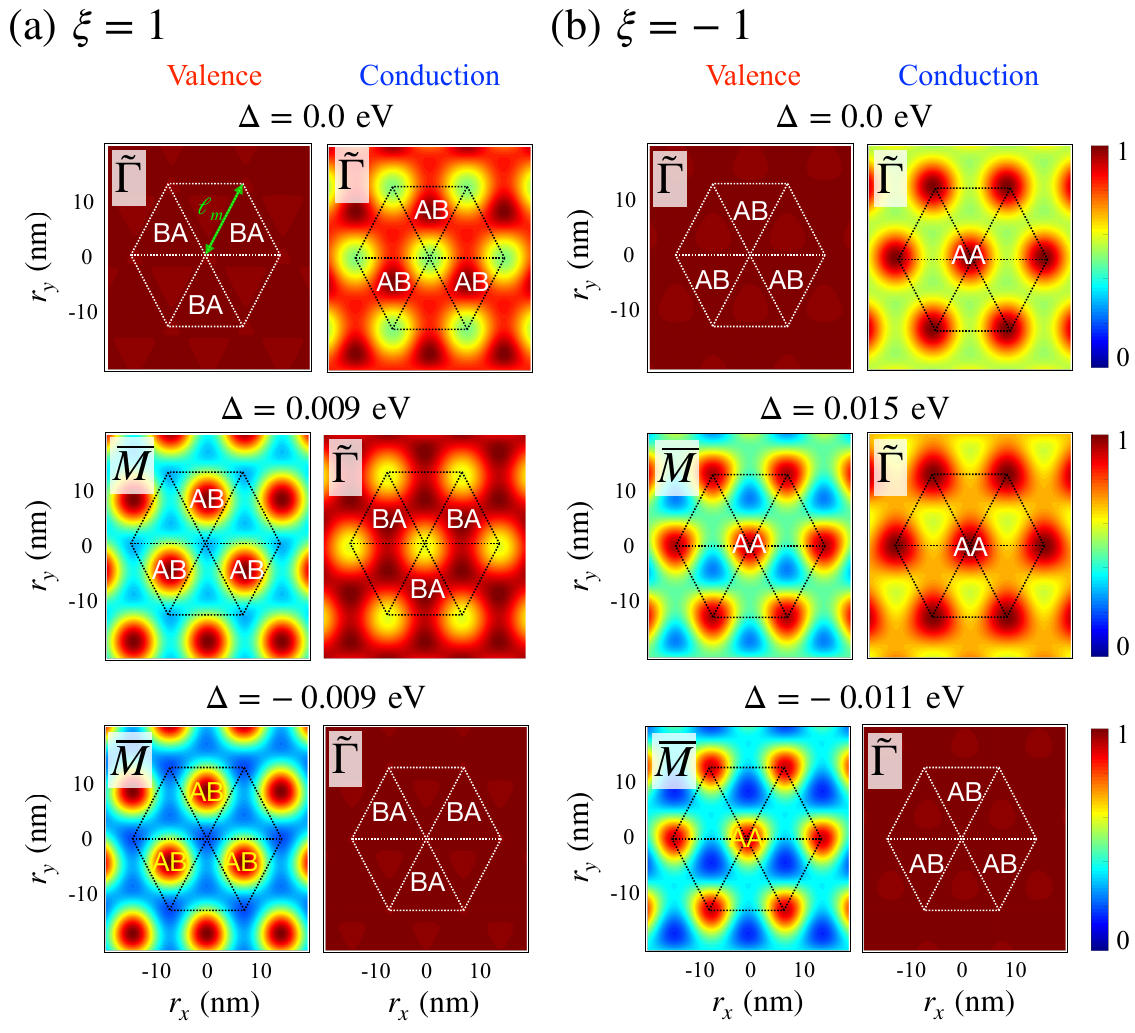} 
\caption{(color online) 
We calculate the $k$-point-projected LDOS $|\psi_{V1(C1)}({\bm k}:{\bm r})|^2$ at twist angle $\theta=0^{\circ}$ and high-symmetry points $\tilde{K}$, $\tilde{K}^{\prime}$, $\tilde{\Gamma}$, and $\overline{M}$, for two different hBN substrate orientations: (a) $\xi = 1$ and (b) $\xi = -1$. We normalize the values by the maximum value, and define $\overline{M}$ as $(\tilde{M}+\tilde{M}^{\prime} + \tilde{M}^{\prime\prime})/3$.
We consider selected interlayer potential differences $\Delta = 0.0, 0.009, -0.009~\rm eV$ for $\xi =1$, and $\Delta = 0.0, 0.015, -0.011~\rm eV$ for $\xi = -1$, and plot the LDOS for a given $k$ point that contributes significantly to the LDOS, as labeled in the upper left corner of each figure.
Our results reveal that the LDOS projected onto $\tilde{K}$, $\tilde{K}^{\prime}$, and $\overline{M}$ tends to localize in one of the local stackings, AA, AB, or BA. However, the $\tilde{\Gamma}$-projected LDOS, especially associated with the $B_4$ band, shows delocalization for zero $\Delta$ in the valence bands and negative $\Delta$ in the conduction bands, regardless of the hBN substrate orientation $\xi=\pm 1$.
}
\label{Fig:LDOS_kpoints}
\end{figure}
In the other substrate orientation, $\xi = -1$, (Fig.~\ref{Fig:LDOS}(b)) with $\Delta = 0$, the LDOS slightly(strongly) concentrate at AA stacking for valence(conduction) band.
Unlike the $\xi=1$ orientation, a positive $\Delta =  0.015$~eV makes the valence band LDOS further spread over all stacking configurations while the conduction band LDOS from AA stacking concentrates at AB stacking strongly.
A negative $\Delta = -0.011$~eV shifts the conduction band states towards the top-most layer, especially at $B_4$, delocalizing the conduction band LDOS,
while the valence band LDOS concentrating at $A_1$ in the bottom layer is strongly localized at AA.

A twist angle tilts the moire unit cell and affects the localization of the valence and conduction band LDOS in 4LG/BN.
For $\xi=1$ at $\theta = 0.54^{\circ}$ (Fig.~\ref{Fig:LDOS}(c)),
similar to the zero twist angle system, the valence band LDOS spreads over the moire unit cell and the conduction band LDOS resides at AB and BA slightly favoring AB stacking for $\Delta=0,~0.009~$eV, while the valence band LDOS resides on AB stacking sites and conduction band LDOS spreads within the moire unit cell for $\Delta=-0.009~$eV.
A positive $\Delta=0.009$~eV shifts the localization center of the conduction band LDOS from BA to the AB.
However, for $\xi=-1$ at $\theta = 0.54^{\circ}$ (Fig.~\ref{Fig:LDOS}(d)),
the carrier densities of valence and conduction bands move to the AA stacking site regardless of the electric fields ($\Delta = 0, 0.015, -0.011$~eV) where the conduction band LDOS at the negative $\Delta=-0.011$~eV spreads almost uniformly over the moire unit cell like in the other hBN substrate orientation and twist angles.

\par In the remaining part of this subsection, we further discuss the delocalization in the isolated nearly flat bands of 4LG/BN, that has some analogies to double bilayer graphene~\cite{Crommie2020}. 
We found that the selective sublattice and $k$-point-projected LDOS at the vHS is helfpul for the study of the delocalization in 4LG/BN. 
From the sublattice-projected LDOS (Fig.~\ref{Fig:LDOS_sublatt}), we found that most of the low-energy states are associated with the $A_1$ site in the bottom-most graphene layer and the $B_4$ site in the top-most layer, and the states from $B_4$ are most important for the flat bands' delocalization where the conduction or valence band LDOS almost equally spreads over all stacking configurations. 
Also, from the layer-dependent sublattice-projected LDOS calculation of 3LG/BN, 4LG/BN, and 5LG/BN (See Fig.~\ref{Fig:LDOS_5layer}),
we verified that the delocalization of the states are associated with the $B_n$ site located far away from the hBN substrate but the degree of localization shows a non-monotonic behavior as $n$ increases in $n$LG/BN for both $\xi = \pm1$. 
When we analyzed the $k$-point-projected LDOS we found that the states from the $\Gamma$ are the most delocalized. 
In Fig.~\ref{Fig:LDOS_kpoints}, we illustrate the $k$-point-projected LDOS at the specific $\bm k$ points, $\tilde{\Gamma}$ or $\tilde{M}$, where the vHS states dominantly come from. Here, $\bar{M}$ is defined as $\bar{M} = (\tilde{M}+\tilde{M}'+\tilde{M}'')/3$ and we calculate $|\psi({\bm k}:{\bm r})|^2$ in Eq.~(\ref{Eq:LDOS}) as the $k$-point-projected LDOS by extracting the wave function at the select $k$-points, ${\bm k}=\tilde{K}, \tilde{K}^{\prime}, \tilde{\Gamma}, $or $\tilde{M}^{(\prime,\prime\prime)}$. 
The $\tilde{\Gamma}$-projected LDOS, especially associated to the $B_4$ site, show delocalization for zero(negative) $\Delta$ in the valence(conduction) bands regardless of the hBN substrate orientation $\xi=\pm 1$, while the LDOS contribution from the other symmetric $k$-points $\tilde{K}$, $\tilde{K}^{\prime}$, and $\tilde{M}^{(\prime, \prime\prime)}$ are localized at either AA, AB, or BA.

\section{Summary and Conclusions}
In summary, we investigated the parameters giving rise to isolated nearly flat bands in rhombohedral four-layer aligned to boron nitride (4LG/BN) and $n$LG/BN multilayer systems using full-bands continuum models. 
We found that 4LG/BN ($\xi = 1$) and 4LG/NB ($\xi = -1$) alignments with zero degree twists have naturally narrow low energy bandwidths (12-20 meV), and further explored the parameter space of interlayer potential difference ($\Delta$) and twist angles ($\theta$).
The band widths of the 4LG/BN are generally smaller than the estimated on-site Coulomb repulsion energy $U\approx25~$meV for a large parameter space up to $\Delta = \pm$~0.1 eV and $\theta = 0\sim0.8^{\circ}$, where $U/W>1$ indicates regions of possible onset of Coulomb ordered phases in 4LG/BN. 
Very narrow band widths ($<5$~meV) are possible with appropriate $\Delta$ values for a range of twist angles $\theta=0\sim0.5^{\circ}$ for both $\xi = \pm1$. 

We also found that isolated flat bands are achievable for specific ranges of $\theta$ and $\Delta$ where both primary ($\delta_p$) and secondary ($\delta_s$) gaps are open. 
In the narrow band regime ($U_{\rm eff}/W \gg 1$), we estimated the screened on-site Coulomb repulsion energy $U_{\rm eff} = U ~e^{-{\ell_m}/\lambda_{D}}$, taking into account the screening effect due to band overlap. As a result, we observe strong particle-hole asymmetric behavior in the 4LG/BN narrow bands. Narrow valence bands for the $\xi=1$ substrate orientation can be found for a large range of parameters $\Delta\approx-50\sim10$~meV and $\theta = 0\sim0.8^{\circ}$, while the conduction bands flatten only near $\Delta\approx 7\sim27$~meV and $\theta\approx 0.5^{\circ}$. 
In contrast, the $\xi = -1$ alignment has a larger parameter space for conduction narrow band regime than the valence narrow band regime. Conduction narrow bands can exist for $\Delta=0\sim40$~meV and $\theta=0\sim0.8^{\circ}$, while valence narrow bands are only possible near $\Delta\approx20\sim25$~meV and $\theta\approx0.5\sim0.6^{\circ}$.
We found that the maximum value of $U_{\rm eff}/W $ occurs when $\theta=0.54^{\circ}$, in the valence band with $\xi =1$ alignment taking $U_{\rm eff}/W = 3.98$, and in the conduction band with $\xi = -1$ alignment giving rise to $U_{\rm eff}/W = 4.7$. Therefore, we suggest that the twist angle of $\theta\approx 0.5^{\circ}$ for the 4LG/BN device has the chance of exhibiting Coulomb-induced ordered states for a wider range of perpendicular electric fields.

Furthermore, we showed that the isolated valence and conduction flat bands have well-defined valley Chern numbers of $C_{V1}^{\nu=\pm 1}=\pm4$ for both $\xi = \pm1$ and $C_{C1}^{\nu=\pm 1}=\pm1(0)$ for $\xi=1$($\xi=-1$) 
even without any external electric fields. 
The valence band valley Chern number remains $C_{V1}^{\nu=\pm1}=\pm4$ for both $\xi = \pm1$ orientations 
for $\Delta=0\sim0.035$~eV and similarly for the conduction bands we have $C_{C1}^{\nu=\pm1}=\pm1$ for $\xi=1$ for $\Delta=0\sim0.010$~eV.
By increasing $|\Delta|<100$~meV, the valence and conduction bands can have various topological states ($C_{V1}= \pm1,\pm2,4$, $C_{C1}=1, 4$). 
Also, we found in $n$LG/BN (3$\sim$8 layer graphene) that the valence band valley Chern number of $n$LG/BN is the same as the number of graphene layers ($C_{C1}=n$) for both $\xi=\pm1$ alignments when $\Delta=0$~eV and $\theta=0^{\circ}$.   

We have then investigated the local density of states associated with the van Hove singularities (vHS) of the 4LG/BN flat bands. In the absence of an interlayer potential difference and twist angle, the valence band states at the vHS are spread over all local stackings, slightly favoring the AB stacking for $\xi = 1$ and AA stacking for $\xi=-1$. Meanwhile, the conduction band states for $\xi = -1$ are strongly localized at AA stacking, while for the $\xi=1$ substrate orientation, they reside mainly at the AB and BA stackings, with the AB stacking being slightly favored. The position and degree of localization vary with the substrate orientation, interlayer potential difference, and twist angle. A negative $\Delta = -0.009$~eV for $\xi = 1$ and $\Delta = -0.011$~eV for $\xi = -1$ enhances the localization strength of the valence band vHS states without moving the localization center, while a positive $\Delta = 0.009$~eV for $\xi = 1$ and $\Delta = 0.015$~eV for $\xi = -1$ move the localization center from AB (AA) to BA (AB) for $\xi = 1$ ($\xi = -1$) substrate orientation.

We can modify the localization of the 4LG/BN nearly flat band states with an electric field. Through sublattice-projected LDOS calculations, we confirmed that the states at the vHS mostly come from the low energy non-dimer sublattice sites, specifically $A_1$ from the bottom-most graphene layer contacting hBN and $B_4$ from the top-most layer. The delocalized states are related with the $B_4$ sites in 4LG/BN.
We examined the LDOS data for other multilayers such as 3LG/BN and 5LG/BN and found that the states populating the farthest $B_n$ site away from the hBN substrate tend to be delocalized at all local stackings, either for the conduction or valence band depending on the sign of $\Delta$.
However, there is seemingly no direct relation between layer number and localization, considering that the $B_3$ in 3LG/BN or $B_5$ in 5LG/BN both show stronger localization compared to 4LG/BN.
Further, we examine the $k$-point projected LDOS in 4LG/BN, which shows that the delocalized $B_4$ state mainly originates from the $\tilde{\Gamma}$ point in the MBZ.

In summary, the 4LG/BN has narrow bands in a large parameter space of twist angles and interlayer potential differences, has various associated valley Chern bands, and the localization of the wave functions can be easily controlled by applying relatively small interlayer potential differences of $|\Delta|<0.1$~eV. The low energy states in 4LG/BN are particularly delocalized compared to other layer numbers that will impact the ground states when we explicitly account for Coulomb interactions. 
Given the high tunability of the system and narrow bandwidths as small as 10~meV we believe that the 4LG/BN and larger $n$ multilayer $n$LG/BN are excellent systems to study flat band physics provided that the hurdles for preparing rhombohedral graphene multilayers can be overcome.

\section{Acknowledgments.}
This work was supported by 
the National Research Foundation of Korea (NRF) 
with grant number 
No.~NRF-2021R1A3A13045898 for Y. P.,
No.~NRF-2020R1A2C3009142 for Y. K., 
No.~NRF-2020R1A5A1016518 for J. J., and
the SERB with grant no. SRG/2022/001102
for B. L. C..
We acknowledge computational support from KISTI Grant No. KSC-2022-CRE-0514 and by the resources of Urban Big data and AI Institute (UBAI) at UOS. J.J. also acknowledges support by the Korean Ministry of Land, Infrastructure and Transport(MOLIT) from the Innovative Talent Education Program for Smart Cities.
\bibliography{myref.bib}
\bibliographystyle{apsrev_nourl}
\begin{widetext}
\end{widetext}

\clearpage
\pagebreak

\begin{appendix}

\renewcommand\thefigure{A\arabic{figure}}  
\setcounter{figure}{0}  

\section*{Appendix}
\section*{A1. F1G0 tight binding model for multilayer graphene}
\label{appendix_dftbands} 
\begin{figure}[!htb]
\includegraphics[width=7.5cm,angle=0]{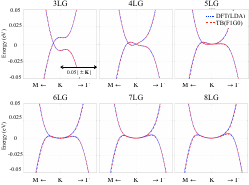} 
\caption{(color online) 
We compare the F1G0 tight-binding model bands (TB(F1G0)) to those obtained from first-principles calculations within the local density approximation (DFT(LDA))  for 3LG-8LG along the $k$ path from M to K to $\Gamma$. The bands from both methods are in good agreement in the energy range of $-0.05\sim0.05~$eV and the $k$-space range of $|{\bm k}|<0.05\left(\frac{4 \pi}{3 a}\right)$ near the Dirac points ${\bm K}_{\nu=1}$.
}
\label{Fig:DFT_compare}
\end{figure}

The effective tight-binding F1G0 model~\cite{accuratemonolayer} utilized in this study is based on grouping the hopping amplitudes by the hopping distances, where the effective Hamiltonian only includes the structure factor functions of $g_0 ({\bm k})$ for the zero distance intra-sublattice hopping and $f_1({\bm k})$ for the first nearest neighbor inter-sublattice hopping. To construct the ABC stacked rhombohedral multilayer graphene Hamiltonian, we adopted the trilayer graphene (3LG) hopping parameters
since they provide a highly accurate tight-binding fit to the first-principles calculations bandstructure in the low energy range near the charge neutrality point (CNP). 
The resulting band structures of the F1G0 tight-binding model for multilayer graphene, as shown in Fig.~\ref{Fig:DFT_compare}, are in good agreement with those obtained from density functional theory (DFT) calculations, within the energy range of $|E|<0.05$eV and the $k$ vector range of $|{\bm k}| < 0.05~\left(\frac{4 \pi}{3 a_G}\right)$ measured from the Dirac points ${\bm K}_{\nu=\pm1}$. We also extended the effective tight-binding model Hamiltonian to 4LG, 5LG, 6LG, 7LG, and 8LG by modifying the adopted 3LG hopping parameters.
To obtain the DFT band structures, we used $Quantum\ ESPRESSO$ with local density approximation (LDA) parametrization (C.pz-rrkjus.UPF), where the graphene lattice constant and the interlayer separation are set to $a = 2.46\AA$ and $c=3.35\AA$. A 60 Ry energy cutoff and a $30\times30\times1$ Monkhorst-Pack $k$ point grid were used.

\section*{A2. Bandwidth in multilayer graphene boron-nitride moire superlattices ($n$LG/BN)}
\label{appendix_multigraphene}

Our investigation revealed that the low-energy valence and conduction bands in 4LG/BN exhibit significantly flatter bandwidths compared to those in 3LG/BN~\cite{chittari}. To gain deeper insights into this trend, we examined the bandwidths for $n$LG/BN where $n$ varies from 3 to 8. For this purpose, we added the effective moiré potential $H^{\xi}_M({\bm k}:{\bm r})$ described in Eq.(\ref{Eq:moire}) to the F1G0 Hamiltonian of $n$LG, as detailed in Appendix A1.
Our analysis is summarized in Fig.\ref{Fig:BS_5layer}, where we show (a)-(b) the band structures of 3LG/BN, 4LG/BN, and 5LG/BN and (c)-(d) the bandwidths of the low-energy valence and conduction bands as a function of the number of graphene layers for both orientations of the BN substrate ($\xi = 1$). 
We observed a nonzero valley Chern number $C^+_{V1} = +n$ for the valence bands in $n$LG/BN for both $\xi=\pm1$, matching with the number of graphene layers. Meanwhile, the conduction bands for $\xi = 1$ have a nonzero valley Chern number $C^+_{C1} = +2$ for 3LG/BN and $C^+_{C1} = +1$ for 4LG/BN and 5LG/BN, while $C^+_{C1} =0$ for $\xi = -1$.
Our analysis also revealed that the bandwidths of the valence and conduction bands decrease from $W\approx0.03\sim0.04~$eV (3LG/BN) to $W\approx0.005\sim 0.010$ eV (5LG/BN) and remain unchanged for further increases in the number of graphene layers up to eight (8LG/BN), as shown in Fig.\ref{Fig:BS_5layer} (c)-(d). Among the $n$LG/BN moiré superlattices, the 5LG/BN with a narrow bandwidth of $W\approx0.005\sim 0.010$ eV is the most attractive candidate, but the 4LG/BN with a bandwidth of $W\approx0.01$ eV is still one of the most attractive candidates. This is because, as the number of layers increases, it becomes more challenging to prepare the samples.

\begin{figure*}[!htb]
\includegraphics[width=17.5cm,angle=0]{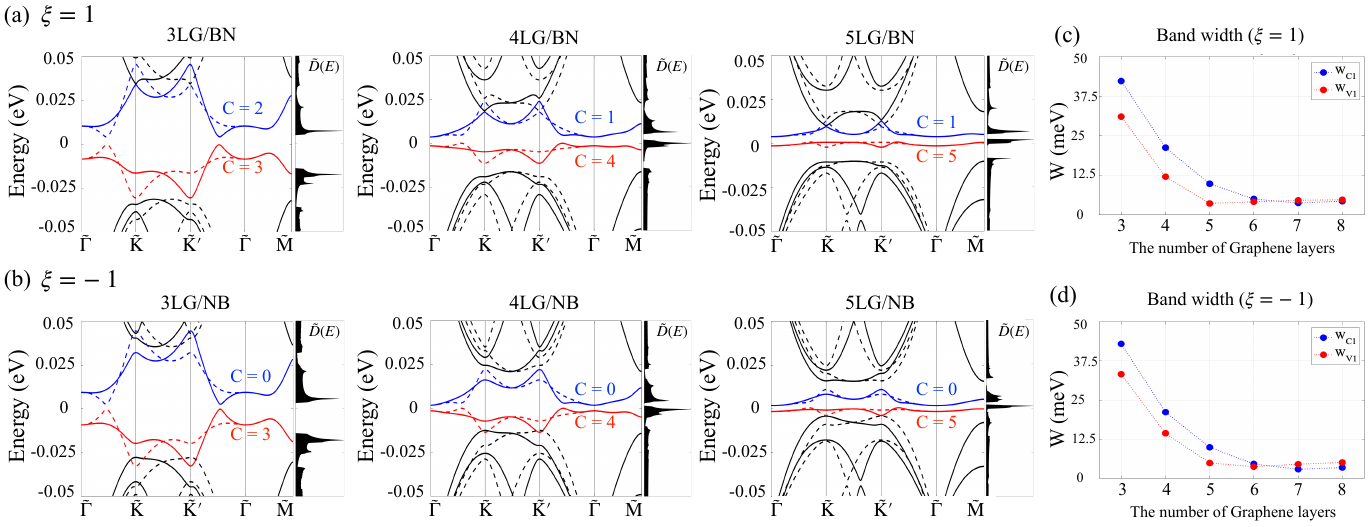} 
\caption{(color online) 
We present the band structure of $n$LG/BN, where $n = 3\sim5$ layers of graphene on (a) $\xi= 1$ and (b) $\xi = -1$ hBN substrates. The band structures for two valleys are shown with solid ($\nu = 1$) and dashed ($\nu=-1$) lines. We observe that the valley Chern number for the $n$LG/BN valence band matches the number of layers, $C^{\nu=\pm1}_{V1} = \pm n$, while those for the conduction bands are $C_{C1}=2$ for the 3LG/BN and $C_{C1}=1$ for the 4LG/BN and 5LG/BN. To investigate the trend in bandwidths of low energy valence and conduction bands, we plot the bandwidths as a function of the number of graphene layers for $n$LG/BN, where $n = 3\sim8$. In (c) and (d), we show the bandwidths for $\xi= 1$ and $\xi = -1$, respectively. We observe that the bandwidths of the valence and conduction bands decrease from $W \approx 40~$meV (3LG/BN) to a minimum value of $W \approx 5~$meV (5LG/BN) for $\xi = \pm1$. The bandwidths remain constant when further increasing the number of graphene layers for $n$LG/BN, where $n = 6\sim8$.
}
\label{Fig:BS_5layer}
\end{figure*}
\begin{figure*}[!htb]
\includegraphics[width=17.5cm,angle=0]{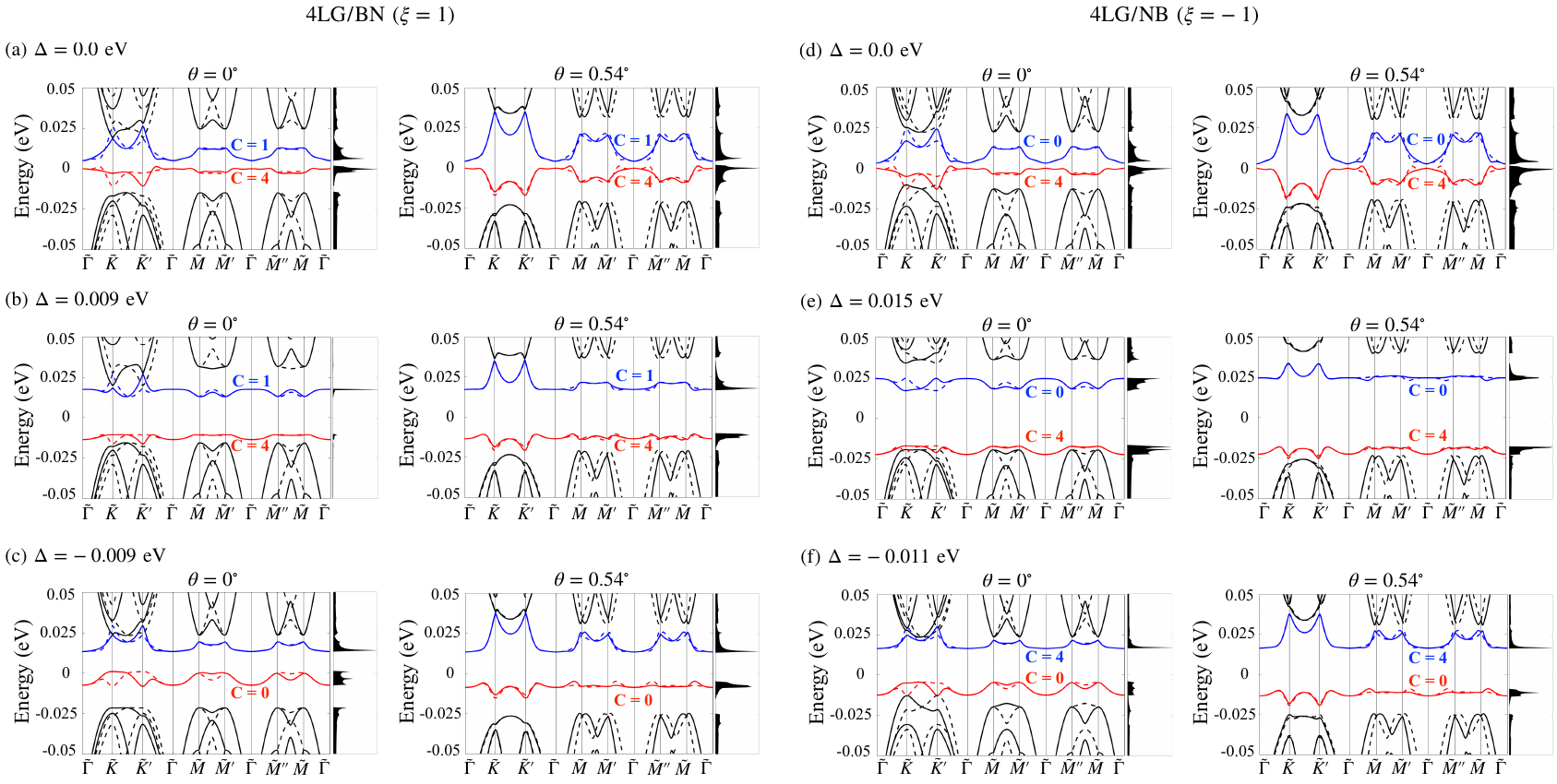} 
\caption{(color online) 
We investigate the impact of interlayer potential difference $\Delta$ on the band structure of 4LG/BN for the two hBN substrate orientations ($\xi = \pm1$). We consider two twist angles (left: $\theta = 0^{\circ}$, right: $\theta = 0.54^{\circ}$) and three values of $\Delta$ for each orientation: (a)-(c) $\Delta = 0.0$~eV, $\Delta = 0.009$~eV, and $\Delta = -0.009$~eV for $\xi = 1$, and (d)-(f) $\Delta = 0.0$~eV, $\Delta = 0.015$~eV, and $\Delta = -0.011$~eV for $\xi = -1$.
We plot the band structure for both valleys, denoted by solid/dashed lines for $\nu=1$ and $\nu=-1$. Our results show that the bandwidth of the valence band decreases with increasing twist angle and interlayer potential difference, while the valley Chern numbers for each low energy valence (conduction) band are unaffected by the twist angle at a given interlayer potential difference.
}
\label{Fig:BS_0deg}
\end{figure*} 
\begin{figure*}[!htb]
\includegraphics[width=17.5cm,angle=0]{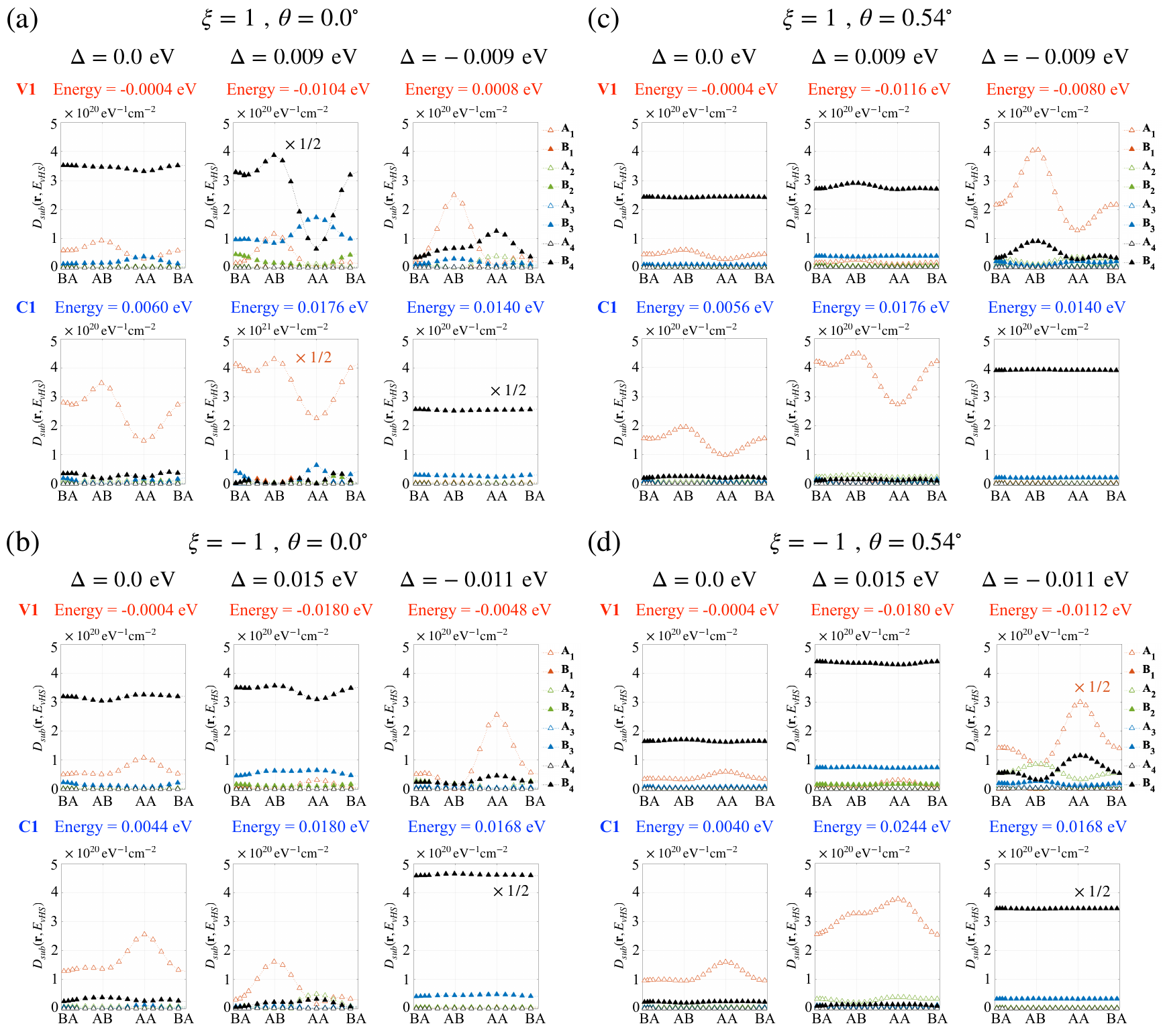} 
\caption{(color online) 
We present the sublattice-projected local density of states (LDOS) of 4LG/BN for various interlayer potential differences ($\Delta$) at the van Hove singularity (vHS) of low-energy valence and conduction flat bands. Specifically, we consider (a) $\Delta = 0.0,0.009, -0.009~$eV for $\xi=1$, and (b) $\Delta = 0.0,0.015, -0.011~$eV for $\xi = -1$, at $\theta=0^\circ$. We also show the corresponding results for $\theta=0.54^\circ$ in panels (c) and (d).
Our analysis reveals that the dominant states in the LDOS stem from $B_4$ and $A_1$ sublattice sites, as denoted by the black-filled and red-open triangle symbols, respectively. Furthermore, the dominant $A_1$ states are primarily localized at AB (AA) stacking sites for $\xi=1$ ($\xi=-1$), while the $B_4$ states spread within the moire unit cell.
}
\label{Fig:LDOS_sublatt}
\end{figure*}

\section*{A3. Effects of a twist angle}

Our study demonstrates that a small twist angle $\theta =~ 0.54^{\circ}$ between graphene and boron nitride has a significant impact on the 4LG/BN band structure. Specifically, we observe that this twist angle leads to the opening of a secondary band gap for both conduction and valence bands, compared to the $\theta = 0^{\circ}$ case. These changes are facilitated by the presence of finite interlayer potential differences.
To further illustrate the effect of the twist angle and interlayer potential differences, we present the band structures of 4LG/BN in Fig.~\ref{Fig:BS_0deg} for twist angles of $\theta = 0^{\circ}$ and $\theta = 0.54^{\circ}$, with select interlayer potential differences of $\Delta = 0, 0.009, -0.009$ eV for $\xi = 1$ and $\Delta = 0, 0.015, -0.011$ eV for $\xi = -1$. 
Notably, we find that the conduction (valence) band isolation for $\xi=1$ ($\xi=-1$) is present at $\theta = 0.54^{\circ}$ in the presence of a small positive (negative) $\Delta$, which is difficult to achieve at $\theta = 0^{\circ}$. This underscores the critical role of the small finite twist angle in achieving band isolation remaining the band flatness.
In addition, we discover that the twist angle $\theta=0.54^{\circ}$ reduces the asymmetry between the two minivalleys at $\tilde{K}$ and $\tilde{K}^{\prime}$ without modifying their topological phases. This adds another intriguing feature to the impact of the twist angle on the electronic properties of 4LG/BN.

\begin{figure*}[!htb]
\includegraphics[width=17.5cm,angle=0]{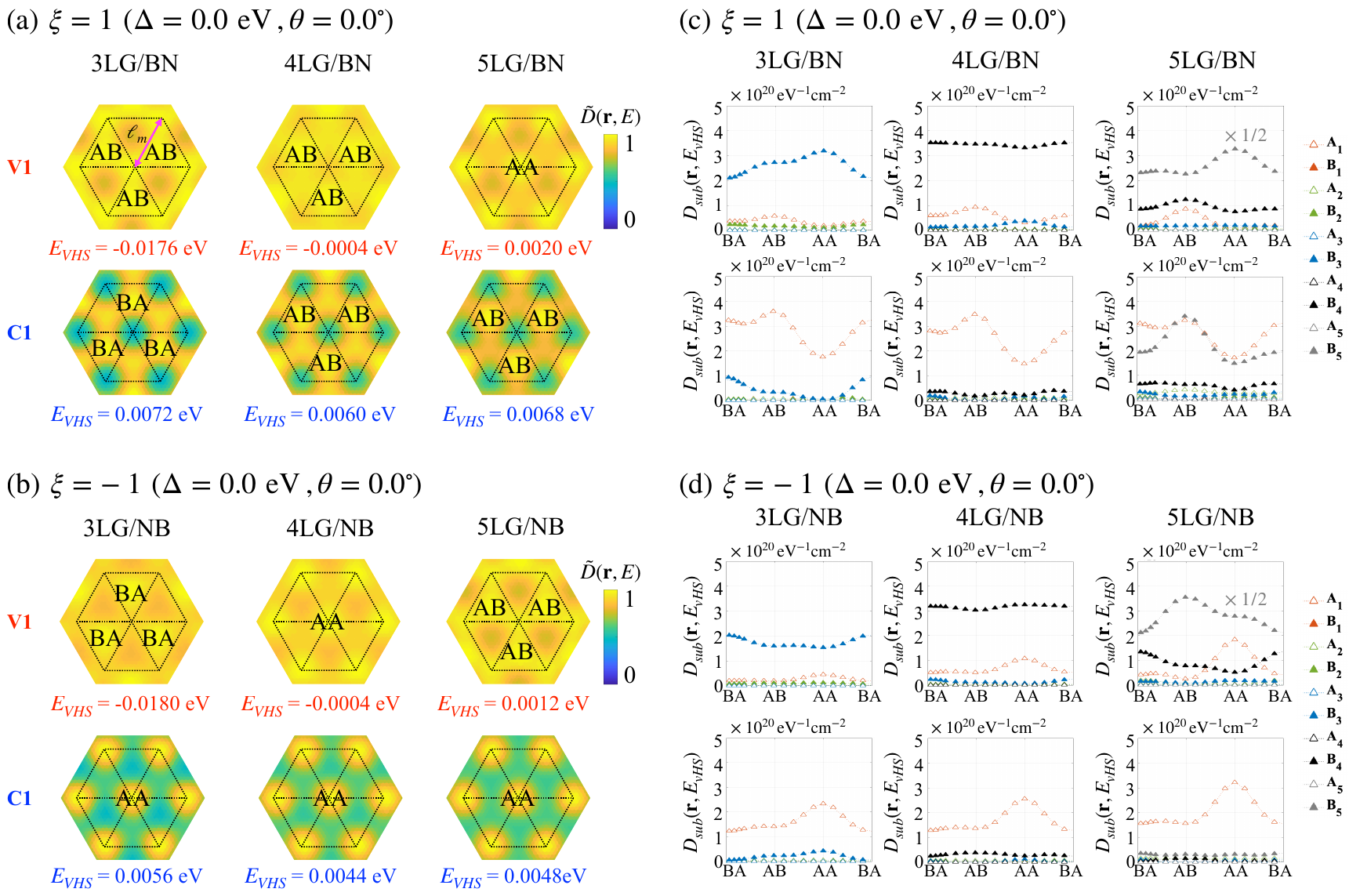} 
\caption{(color online) 
We present the normalized local density of states (LDOS) at the van Hove singularity (vHS) of the conduction (C1) and valence (V1) bands for 3LG/BN, 4LG/BN, and 5LG/BN for two different hBN orientations: (a) $\xi=1$ and (b) $\xi=-1$. Interestingly, we find no clear trend in the localization of states with increasing number of graphene layers. Additionally, we compare the sublattice-projected LDOS of these systems calculated at the vHS of flat bands for the (c) $\xi=1$ and (d) $\xi=-1$ hBN orientations. Our analysis reveals that the dominant states originate from the low energy $B_n$ and $A_1$ sublattice sites. We find that the $B_n$ states far from the hBN susbstrate exhibit delocalization in the case of $n$LG/BN.
}
\label{Fig:LDOS_5layer}
\end{figure*}

\section*{A4. Sublattice projected local density of states}
\label{appendix_bands}

To understand the delocalization observed in 4LG/BN, we analyzed the sublattice-projected LDOS of the low-energy flatbands. Our results, shown in Fig.~\ref{Fig:LDOS_sublatt}, indicate that the localized states originate from non-dimer low energy sites, such as $A_1$ and $B_4$. In particular, $A_1$ states are well localized at AB(AA) stacking sites for $\xi = +1$($\xi = -1$), while $B_4$ states exhibit delocalization. The population of the sublattice-projected LDOS depends on the interlayer potential difference ($\Delta$), with more $B_4$ states occupying the valence bands for zero and positive $\Delta$, and the conduction bands for negative $\Delta$.
To confirm that the delocalization of $B_4$ states is related to the distance from the hBN substrate, we calculated the sublattice-projected LDOS for 3LG/BN, 4LG/BN, and 5LG/BN, as shown in Fig.~\ref{Fig:LDOS_5layer}. Our analysis reveals that valence band states are spread over the moire unit cell for all systems, with the delocalized states mostly originating from $B_3$, $B_4$, and $B_5$ for 3LG/BN, 4LG/BN, and 5LG/BN, respectively. However, the extent of localization is not linear with the distance from the hBN substrate, as the delocalized states of 4LG/BN are much more spread over the moire unit cell than those of 5LG/BN. This result suggests that the delocalization of $B_4$ states is influenced not only by the distance from the hBN substrate but also by other factors, such as the interlayer potential difference.

\end{appendix}
\end{document}